\pdfoutput=1

\documentclass{JINST}
\let\ifpdf\relax
\usepackage{url}

\title{First operation and performance of a 200~lt double phase
 LAr LEM-TPC with a 40$\times$76~cm$^2$ readout}

\author{A.~Badertscher, A.~Curioni, U.~Degunda,
  L.~Epprecht, A.~Gendotti, S.~Horikawa, L.~Knecht,
  D.~Lussi, G.~Natterer, K.~Nguyen,
  F.~Resnati, A.~Rubbia\thanks{Corresponding
    author.}\, and T.~Viant~\\
  ETH Zurich, Institute for Particle Physics,\\
  CH-8093 Z\"{u}rich, Switzerland\\ 
  E-mail: \email{Andre.Rubbia@cern.ch}}

\abstract{ 
In this paper we describe the design, construction, and operation 
of a first large area double-phase liquid argon Large
Electron Multiplier Time Projection Chamber (LAr LEM-TPC). The detector has a 
maximum drift length of 60~cm and the readout consists of a $40\times
76$~cm$^2$ LEM and 2D projective anode to multiply and collect drifting
charges. Scintillation light is detected by means of cryogenic
PMTs positioned below the cathode. 
To record both charge and light signals, we have developed 
a compact acquisition system, which is scalable up
to ton-scale detectors with thousands of charge readout channels. The acquisition
system, as well as the design and the performance of custom-made charge
sensitive preamplifiers, are described. The complete experimental setup
has been operated for a first time during a
period of four weeks at CERN in the cryostat of the ArDM
experiment, which was equipped with liquid and gas argon purification
systems. The detector, exposed to cosmic rays, recorded
events with a single-channel signal-to-noise ratio in excess of 30
for minimum ionising particles. 
Cosmic muon tracks and their $\delta$-rays were used
to assess the performance of the detector, and to
estimate the liquid argon purity and the gain at different 
amplification fields.
}


\begin{document}

\section{Introduction}
\label{sec:introduction}

The liquid argon time projection chamber (LAr-TPC)~\cite{Amerio:2004} is a
charge imaging detector which allows to reconstruct tracks in
three dimensions as well as the deposited energy. In this context, 
the Giant Liquid Argon Charge Imaging ExpeRiment (GLACIER) is a
concept proposed for a future observatory for neutrino physics and
nucleon decay searches
~\cite{Rubbia:2004tz,Rubbia:2009md,Badertscher:2010sy}, which
could be scalable up to gigantic masses of 100 kton.
The key and innovative feature of the GLACIER design
is the double phase LEM-TPC 
operation mode with adjustable gain and 2D projective 
readout~ \cite{Badertscher:2011sy,Badertscher:2009av,Badertscher:2008rf}.
The ionisation charge is extracted to the
argon gas phase where it is amplified by a Large Electron Multiplier (LEM) which
triggers Townsend multiplication in the high field regions in the
LEM holes. The charge is collected and recorded on a two-dimensional and segmented
anode. This principle has two  
main advantages: 1. the  gain in the LEM is adjustable, i.e. the signal
quality can be optimised, and 2. the signals collected on the two
readout views are unipolar and symmetric which facilitates the event
reconstruction. 
In addition to the ionisation charge, scintillation light is emitted
by excited argon diatomic molecules (excimers). The detection of the
scintillation light is fast, thus providing the event time reference $T_0$. 

We have previously reported on the construction and successful
operation of a $10\times 10$~cm$^2$ LAr LEM TPC with 2D
projective readout~\cite{Badertscher:2011sy}. 
Following a staged approach,
the next step towards large-area detectors was the construction
of the so far largest singular unit of this kind, with a readout
cross-section of $40\times 76$~cm$^2$ (of a $\sim$0.5~m$^2$ scale) and a drift of 60~cm,
for a total volume of about 200~lt. 
The complete double phase readout system, henceforth referred to as
\textit{charge readout sandwich}, consists of two extraction grids, the LEM
and a 2D anode, stacked on top of each other. It is a standalone
structure that can be considered as basic module to cover larger
surfaces of bigger detectors. 
After the production and assembly, the detector was
successfully operated during one month inside the cryostat of 
the ArDM experiment operated on surface at 
CERN~\cite{Badertscher:2012dq,Marchionni:2010fi,Rubbia:2005ge}.

This paper is subdivided in four parts: in Section~\ref{sec:detector}
we first describe the experimental setup and design and construction
of the charge readout sandwich. Section~\ref{sec:readout-electronics} is about the design
and 
the layout of the charge and light acquisition, including a detailed paragraph on the
custom made charge sensitive preamplifiers. Finally, in
Section~\ref{sec:performance} we report on the performance of the detector
and the data acquisition system, which then leads to the conclusions in
Section~\ref{sec:conclusions}. 

\section{The experimental setup}
\label{sec:detector}

\begin{figure}[tb]
  \centering
  \includegraphics[width=0.8\textwidth]{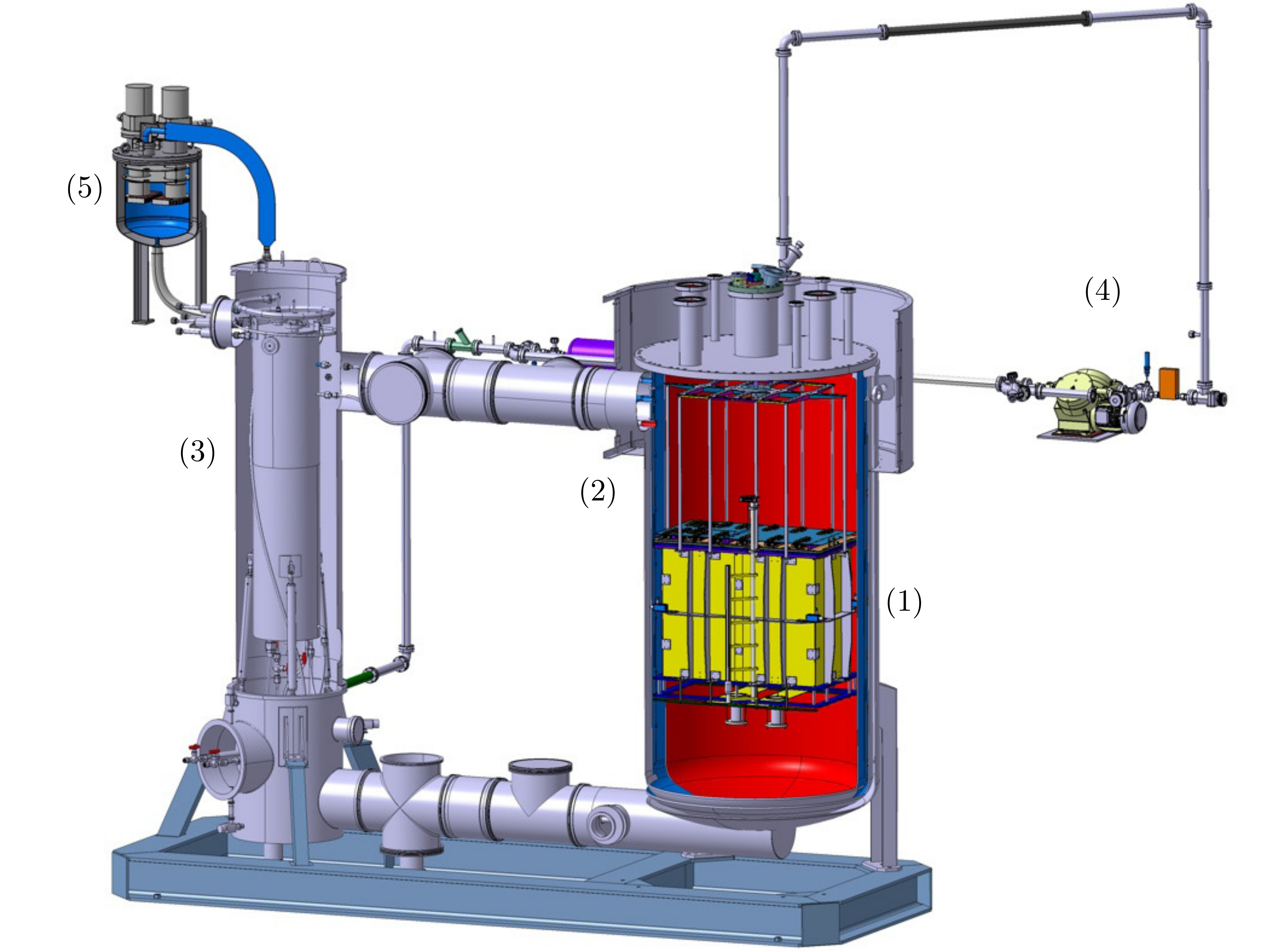}
  \caption{CAD drawing of the system. The detector~(1) hanging from
    the top flange of the ArDM vessel~(2), the liquid argon
    recirculation column~(3) containing the custom made pump and
    filter, the gas recirculation system~(4) and the re-condenser,
    with two cryocoolers~(5).} 
  \label{figure:250LCAD}
\end{figure}

\begin{figure}[t]
  \centering
  \includegraphics[width=1\textwidth]{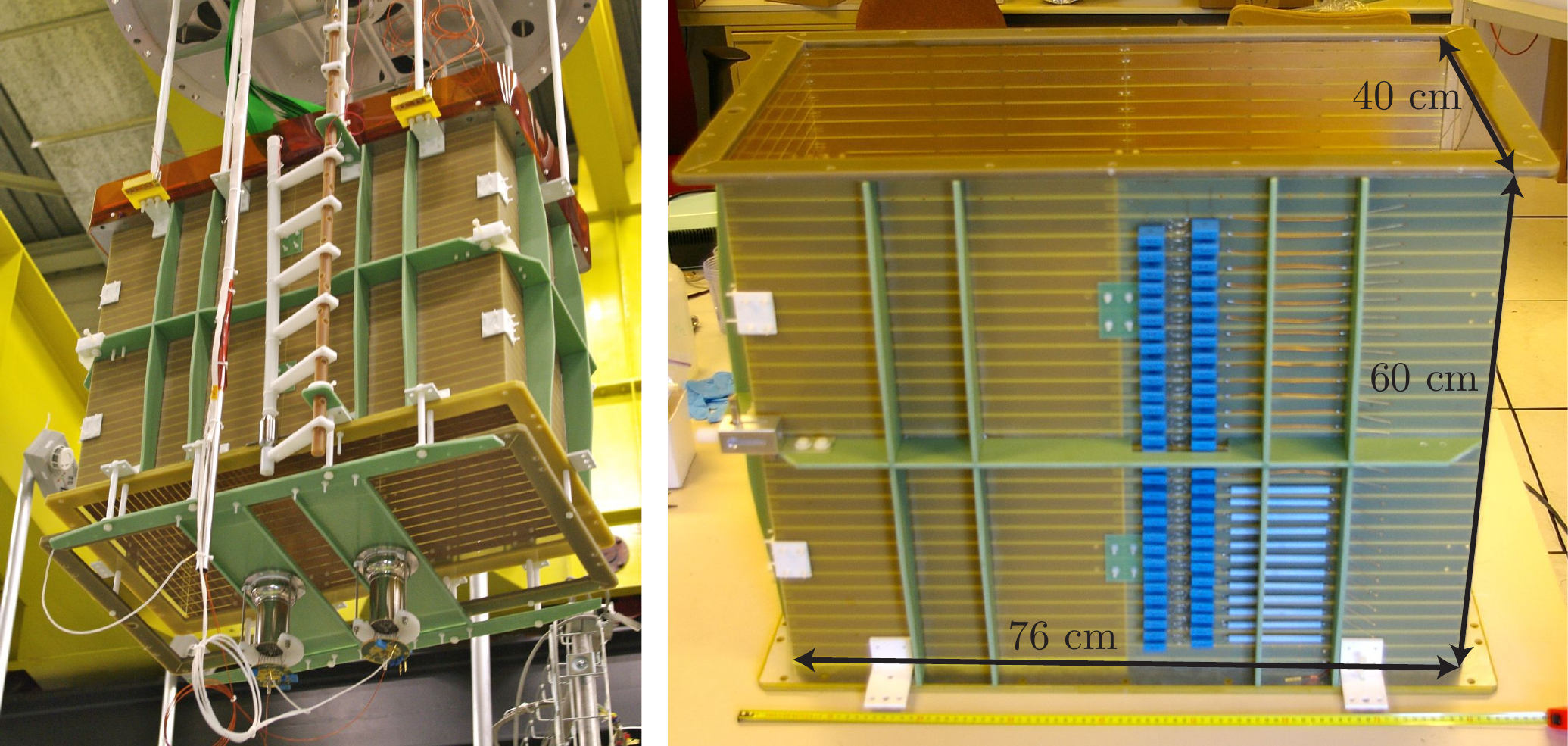}
  \caption{Pictures of the detector. Left: fully assembled detector
    hanging from the top flange of the ArDM vessel.  Right: TPC during
    the assemblage phase.} 
  \label{figure:250L}
\end{figure}

The cryogenic setup developed for the ArDM experiment (See Figure~\ref{figure:250LCAD})
has been employed to provide the required thermodynamic stability 
and liquid argon purity for the
double phase operation of the chamber. 
The ArDM cryogenic infrastructure consists of
two main parts: the detector vessel,
and the liquid argon cryogenic circuit for LAr purification and boil off recondensation.
Both are surrounded by a separate volume that acts as external bath. 
The ArDM vessel is a vacuum tight cylindrical container (1~m in
diameter and about 2~m high) connected to the liquid recirculation column. 
Due to the large volume of liquid argon ($\sim$1~m$^3$) that can be contained
in the ArDM vessel, 
both liquid and gas
purification are implemented. The liquid recirculation, consisting of a
liquid pump and a purification cartridge, is custom made, while
the gas recirculation employs commercial solutions for the pump and the 
cartridge.
The thermodynamic conditions are controlled by two
Gifford-McMahon cryocoolers\footnote{Air-cooled Cryomech AL 300.},
that re-condense the boil off argon of the external bath. The
cryocoolers are controlled by a Programmable Logic Control (PLC) unit,
that also monitors all the slow control processes related to the
pressures and temperatures of the system. The detector is hanging from
the top flange of the vessel, as shown in Figure~\ref{figure:250L}.
Two 3~inch cryogenic PMTs\footnote{Hamamatsu R11065.} are installed
below the cathode, protected by a metallic grid kept at ground. 
The window of one PMT was coated with Tetraphenyl 
Butadiene (TPB) in a polymer matrix in order to convert the
scintillation light into visible light~\cite{Boccone:2009kk}. The
primary scintillation signals are used as trigger for the charge
readout.

\subsection{The LAr LEM-TPC}
\label{sec:theLarLEM-TPC}
\begin{figure}[t]
  \centering
  \includegraphics[width=1.0\textwidth]{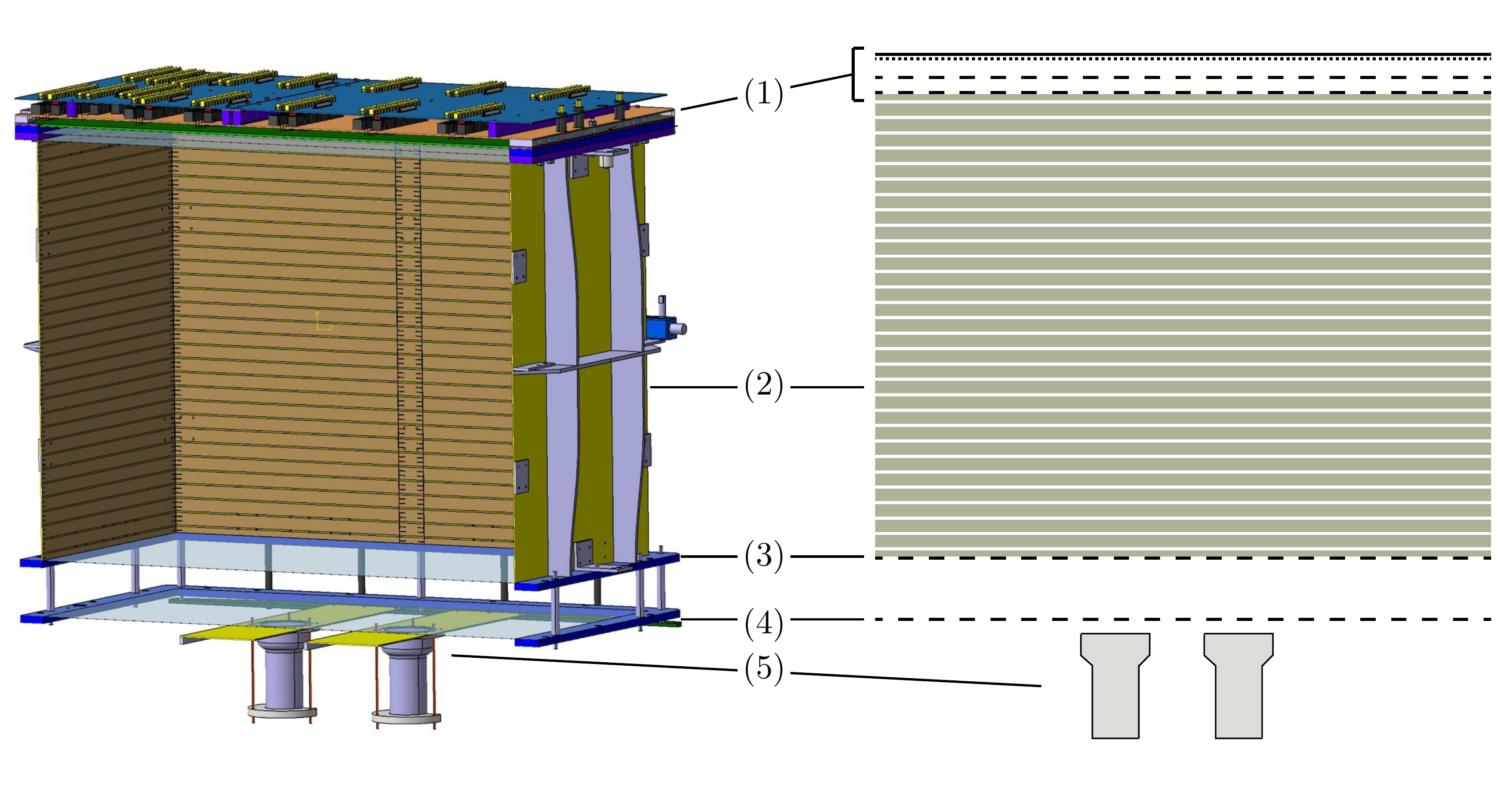}
  \caption{CAD drawing showing a cut through the LEM TPC and the
    detector layout: the charge readout sandwich~(1) embedding extraction
    grids, LEM and 2D anode on top, the PCB drift box with printed copper
    strips (2) below, delimited on the bottom by the cathode grid~(3) and
    an additional grid~(4) to protect the two PMTs~(5) from cathode discharges.} 
  \label{fig:DriftCage}
\end{figure}

The drift cage is made out of Printed-Circuit-Board (PCB) plates. As shown in
Figure~\ref{fig:DriftCage}, it is limited on the bottom 
by the cathode grid and on the top by the extraction grids.
It is 60~cm high and $40 \times 76$~cm$^2$ in cross section to match the
dimensions of the readout sandwich. Thirty-one field shaping electrodes
spaced 2~cm are printed with PCB techniques on the internal surfaces
of the plates. They ensure the uniformity of the drift field along the
entire volume. The high voltage for the drift field is provided by an
immersed 30~stages
Greinacher voltage multiplier circuit, also known as
Cockcroft-Walton voltage multiplier. 
A more detailed description of
the Greinacher voltage multiplier and its performance can be 
found elsewhere~\cite{Badertscher:2012dq,Horikawa:2010bv}.
The components are directly
mounted on the external PCB wall completely immersed in the liquid
(see Figure~\ref{figure:250L}, right). 
The DC ground point of the multiplier is connected to the extraction 
grid in liquid, all the multiplication stages to the corresponding field 
shaping electrodes, finally the DC output (the last stage) to the cathode. 
The last stages are connected to the respective electrodes through
20~M$\Omega$ resistors that limit the current in case the cathode grid
discharges to the wall of the cryostat. The maximum achieved drift
field was 1~kV/cm, with 60 kV output voltage on the cathode. 
The potential of all the stages can be globally
shifted by a DC voltage, in order to match the needed voltage on the
extraction grid in liquid without affecting the drift field. 
Four capacitive level meters are installed around the charge
readout sandwich on top of the detector. The measurement of the liquid level
in different positions allows to adjust the detector with respect to
the liquid level. In fact, the entire vessel can be tilted by means of
screw driven movable stands. 

\subsection{The charge readout sandwich}
\begin{figure*}[t]
\centering
\includegraphics[width=1.0\textwidth]{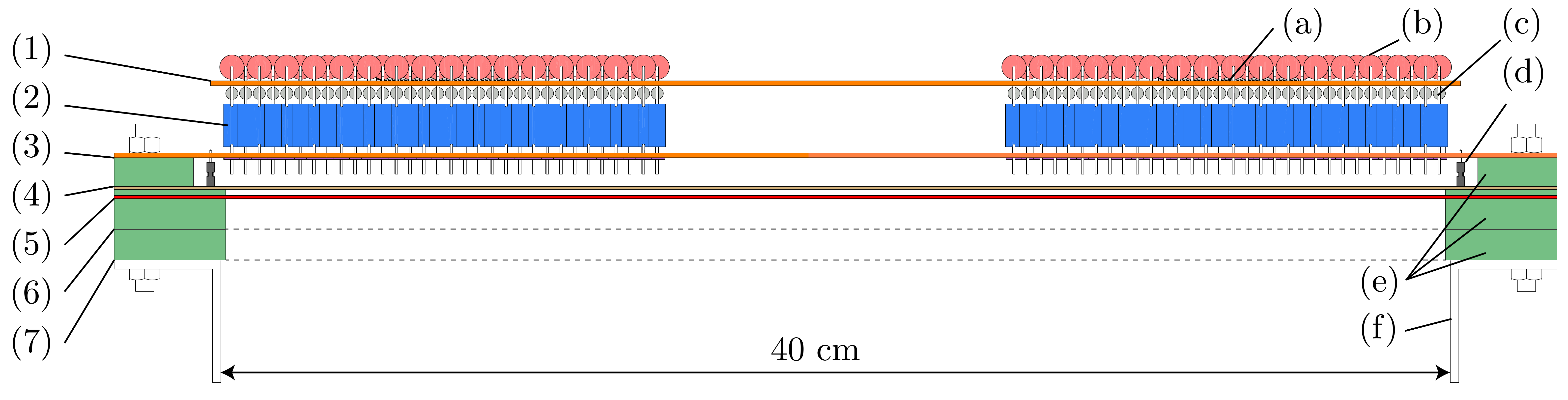}
\caption{2D CAD drawing of the readout sandwich, providing
  charge extraction, amplification and detection. The main elements
  are listed on the left (numbers):
  the signal distribution PCB plane (1),
  the HV decoupling capacitors (2),
  the HV distribution PCB plane (3),
  2D~anode (4),
  LEM (5), 
  extraction grid in gas (6) and in liquid (7).
  Other components shown in the graphics (letters) are 
  the connectors for the signal cables (a) behind
  the surge arresters (b),
  33~$\Omega$ resistors~(c),
  the multi pin connectors between the anode and the HV distribution
  plane~(d),
  epoxy frames and spacers (e)
  and finally supporting angles to attach the
  sandwich to the drift volume (f). 
}
\label{figure:CADsandwich}
\end{figure*}

The readout sandwich, shown schematically in
Figure~\ref{figure:CADsandwich}, consists of the two extraction
grids, the LEM, the 2D projective readout anode, the spacers and the signal
routing PCBs. The design parameters of the LEM are the same as the $10
\times 10$~cm$^2$ prototype operated in a 3~lt
detector~\cite{Badertscher:2011sy}, and the same 
applies for the 2D anode. 

\begin{figure}[t]
  \centering
  \includegraphics[width=1\textwidth]{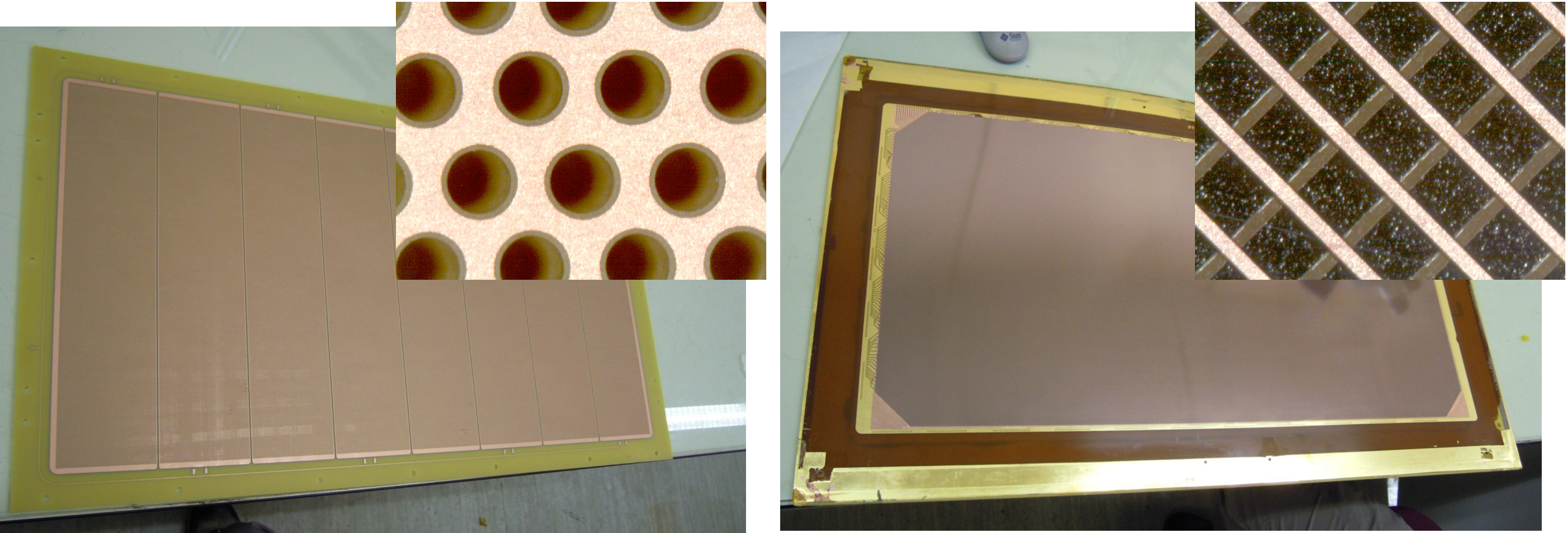}
  \caption{Picture of the $76 \times 40$~cm$^2$ LEM with close-up of
    its holes (left) and picture of the 2D anode with close-up of its strips (right).} 
  \label{figure:AnodeLEM}
\end{figure}

The 1~mm thick LEM has an active area 
of $40 \times 76$~cm$^2$ with about $5 \times 10^5$ holes drilled. 
To limit the charge involved during a discharge, the LEM electrodes
are divided in eight sectors along the long side and powered through
500~M$\Omega$ resistors. When a spark occurs, only the affected sector
is discharged, because the distance between the electrodes of two
sectors (1.6~mm) is enough to avoid the propagation of the discharge
from one sector to the other.
The LEM was produced by a company specialised in PCB manufacturing\footnote{Eltos~S.p.A., San
  Zeno (AR) Italy.}. In order to maximise the amplification gain,
40~$\mu$m thick dielectric rims were chemically etched~\cite{Badertscher:2011sy}. 
To prevent the
surface from corrosion, the metal electrodes were
passivated. A picture of the LEM with a closeup of the holes is shown
in Figure~\ref{figure:AnodeLEM}.
To maintain a uniform distance from the anode, a 2 mm thick spacer is
installed in between the top LEM face and the anode strips. Its shape
matches the division in sectors of the LEM in order not to introduce
additional dead space. The same structure is installed below the LEM to
reduce the sagging due to its weight. 

The anode, produced at the CERN TS/DEM workshop, has the same active
area as the LEM and 256 channels per view, but is not subdivided into sectors. 
In view of a possible test beam, in
order not to have any strip parallel to the incoming particles, the
strips are rotated by 45$^o$ with respect to the long anode side. 
Pictures of the anode and a close view of the strips are shown in
Figure~\ref{figure:AnodeLEM}.

\begin{figure}[t]
  \centering
  \includegraphics[width=1.0\textwidth]{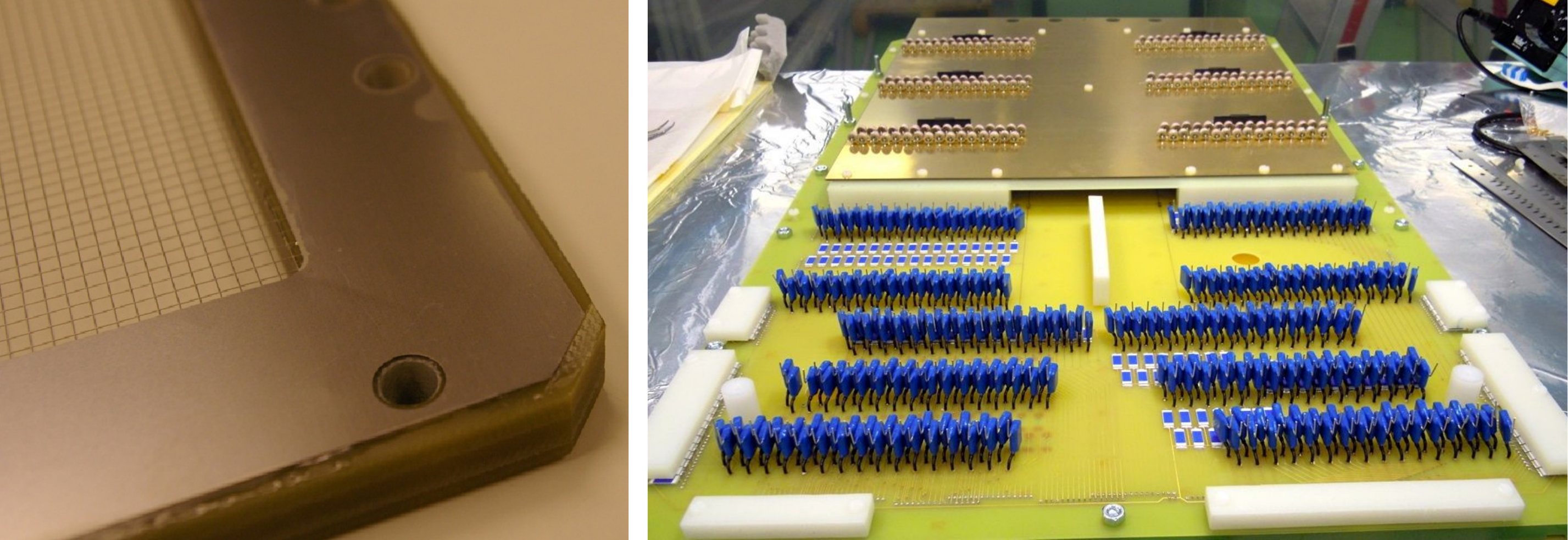}
  \caption{Pictures of the details of the detector: 
  one of the extraction grids (left) and the charge readout 
  sandwich during the assembling (right): the half in the back is already 
  covered with the signal plane, in the foreground the HV distribution plane 
  with the decoupling capacitors and the 500~M$\Omega$ resistors is seen.} 
  \label{figure:GridSignalPlane}
\end{figure}

The two extraction grids, with a distance of 1~cm from each other, are
mounted 1.2~cm below the LEM. Built with the same technique as the
cathode and the PMT protecting grid, they are made of 150~$\mu$m
thick stainless steel foils chemically etched, to leave a square grid
with a pitch of 3~mm and 150~$\mu$m thick \emph{wires}. A picture of
the grid is shown on the left of Figure~\ref{figure:GridSignalPlane}. 
They are glued to 1~cm thick PCB frames, that provide the right
tension to keep them flat. For construction constraints, instead of
stretching the grid, the frame was compressed during the gluing
phase. The alignment between the two grids is within a wire diameter
in the center, but it becomes worse at the edges, because the two
grids are inevitably stretched slightly differently. Potentially, this
can compromise the grid transparency, that can be restored increasing
the field between the top extraction grid and the bottom LEM electrode. 
Image distortions can be caused by the field configuration around the
extraction grids, and for this reason the pitch of the grids was
chosen to be equal to the readout strip pitch. 

The top part of the charge readout sandwich, shown on the right of
Figure~\ref{figure:GridSignalPlane}, hosts the components to 
feed the bias voltages, to decouple the signal from the
high voltage and to protect the electronics from discharges. 
\begin{figure}[t]
  \centering
  \includegraphics[width=1.0\textwidth]{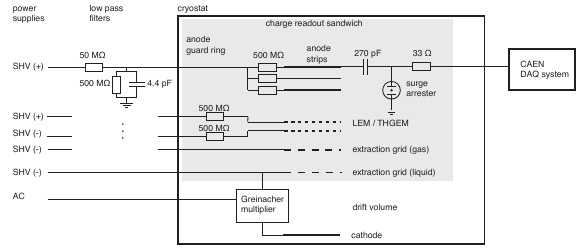}
  \caption{Electrical scheme of the TPC with the HV connections on the
    left and the data acquisition system on the right. The gray box
    shows the components which are embedded in the charge readout
    sandwich.} 
  \label{figure:electricScheme}
\end{figure}
As displayed in the scheme of Figure~\ref{figure:electricScheme}, each
electrode is connected independently via SHV feedthrough and a low pass
filter to a separate channel of the power supply. This configuration
allows to apply arbitrary field configurations. In order to be able to
operate the anode at positive voltages, each of the 512~readout strips
is connected via 500~M$\Omega$ resistors to the
guard ring of the anode.  They are mounted on a first \emph{signal
  routing plane}, which is positioned on top of the anode. The 270~pF
HV decoupling capacitors connect this
first plane to a second one, where a discharge protection circuit is
installed as well as the connectors for the sixteen signal
cables.  Finally, the signal cables are fed through the cryostat and
connected to the charge acquisition system. Both the discharge
protection circuit as well as the readout electronics and acquisition
system are detailed in the following section.

\section{Readout electronics}
\label{sec:readout-electronics}
\begin{figure*}[t]
  \centering
  \includegraphics[width=1.0\textwidth]{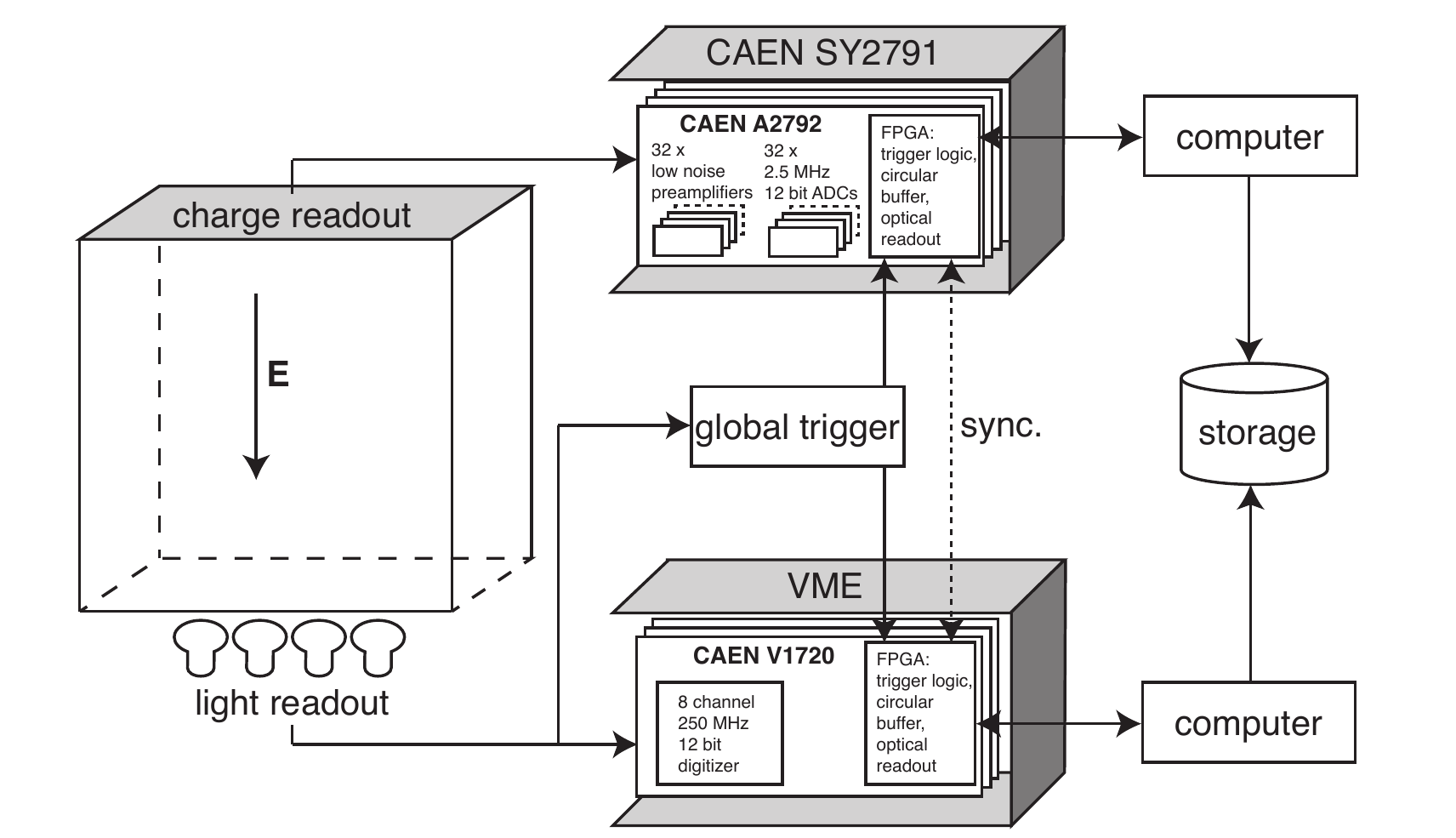}
  \caption{Schematic diagram of charge and light readout organisation
    of a LAr-TPC.}
  \label{fig:intro}
\end{figure*}

A general LAr TPC acquisition layout for charge and light
is presented in Figure~\ref{fig:intro}.  Both acquisition systems can be triggered
by the prompt scintillation light detected with an array of
PMTs immersed in LAr. For the acquisition of the light signals,
commercially available digitizers, such as the 250~MHz V1720 from
CAEN\footnote{CAEN S.p.A., \url{http://www.caen.it}.}  can be used. In order
to acquire the charge data, we developed, in collaboration 
with CAEN, a new charge acquisition system. We chose a scalable and
compact design, which could be used for detectors up to one ton scale
with $O(1000)$ readout channels. As shown in Figure~\ref{fig:intro}, charge
sensitive preamplifiers, ADCs and acquisition logic are placed on
a single board. The purpose of the system is to continuously acquire
simultaneously the waveform of each readout channel, maintaining a
precise time synchronisation between different channels. In addition
we implemented programmable triggers, based on the ionisation charge
digitised signals. We also chose to have preamplifiers directly
pluggable (and thus exchangeable) on the acquisition boards, to be optimised to the needs
of specific detectors. In our case we have developed custom made
preamplifiers for unipolar charge collection signals, with
sensitivities down to a few thousand electrons. 

\subsection{Design requirements}
\label{sec:design-requirements}
The final goal of a LAr TPC detector is to allow the 3D reconstruction
of an ionising particle trajectory crossing the LAr volume and the
determination of the produced ionisation charge along its path. Starting from the
basic interaction of a charged particle with LAr, 
ionisation charge is produced as well as scintillation light (128~nm),
which can be detected with cryogenic PMTs coated with TPB wavelength shifter~\cite{Boccone:2009kk}. The
ionisation electrons, under the action of electric fields typically
between 500~V/cm and 1000~V/cm, are drifted with a speed
between 1.6~mm/$\mu$s and 2~mm/$\mu$s towards the liquid
surface~\cite{Walkowiak:2000wf}. Finally in a LEM-TPC,
after extraction into the gas phase and amplification inside the LEM
holes, the electrons are collected on the electrode strips of a 2D~anode, which are
read out using low-noise charge preamplifiers. 

We have already shown
in~\cite{Badertscher:2011sy} that effective gains 
of about 30 are achievable in stable, long-term conditions.  
The effective gain is
defined
as the ratio of the effectively collected charge on the readout electrodes
compared to the one deposited in the liquid argon medium, however
correcting for potential charge losses due to impurity attachment.
The anode consists of 
two sets of perpendicular readout
strips ($x$ and $y$ coordinate), designed such that each set collects
about half of the incoming charge. Taking into account that the pitch of the
LEM holes is $O$(1~mm), the pitch of the readout strips is set to
3~mm. The drift coordinate $z$ can be directly obtained by measuring the
time difference between the prompt scintillation light used as a
trigger of the event and the collection time of the charge. 
The charge acquisition system has to be able to
continuously acquire simultaneously the waveform of each readout
channel with a precise time synchronisation, over an
acquisition window lasting longer than the maximal drift time
(typically $O$(ms)). In
addition the system has to be compact in order to be placed as close
as possible to the detector. Even with signal cables as short as possible, the
capacitance at the input of the charge sensitive preamplifier
in our setup is about 200~pF. The mean charge per readout strip in our
anode, deposited by a minimum ionising particle (MIP) parallel to the
readout plane and perpendicular to the readout strips, is
 $\sim1.5$~fC/strip without gain from the LEM. 
 For the design of our charge sensitive preamplifier, we hence
required a signal to noise ratio of  $>$15 for 1.5~fC input charge
 and a combined detector and cable capacitance of about
200~pF. With a preamplifier sensitivity of about 10~mV/fC and a 12~bit
ADC with a full range of 3.3~V, the RMS noise is comparable with the
least significant bit of the ADC.

In order to achieve the low noise requirements, a first charge integrating
stage is followed by a signal shaper, consisting of an integrator and
a differentiator. In order to obtain a reconstruction precision of
about 1~mm for the $z$-coordinate, a fast signal rise time of about
0.5~$\mu$s combined with a sampling rate of about 2~MHz are required. For
the falling time constant, a few microseconds is an acceptable
compromise between the signal to noise requirement and the double
track resolution.

\subsection{Charge sensitive preamplifiers}
\label{sec:preamp}
  In this section  we first describe the custom design of our front-end
  charge preamplifier specially optimised for our detector, 
  then we present its performance in terms of pulse
  shaping and noise characterisation for different input capacities.
  
  \begin{figure*}[t]
    \centering
    \includegraphics[width=1.0\textwidth]{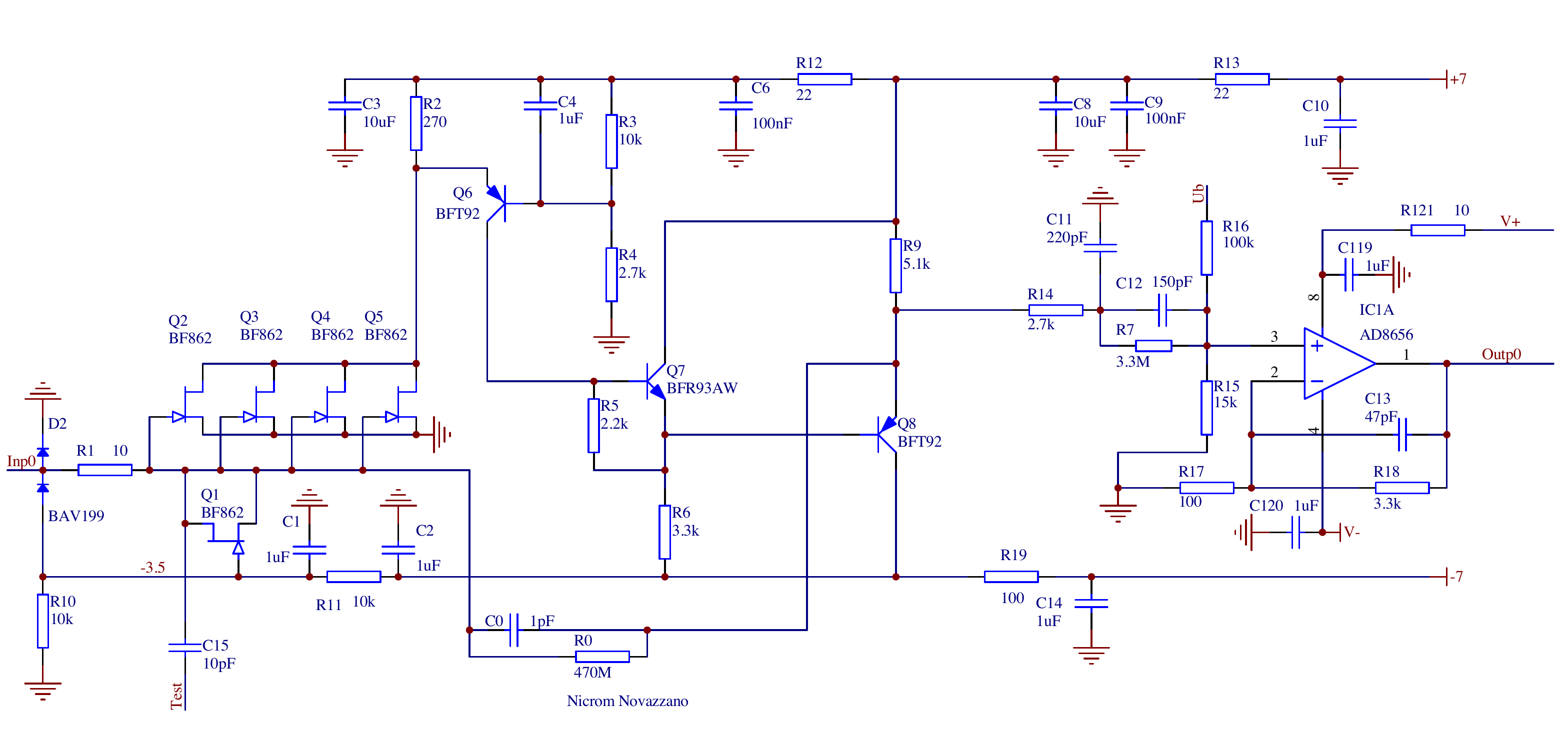}
    \caption{Schematic diagram of the charge sensitive preamplifier.}
    \label{fig:preamp-scheme}
  \end{figure*}

\begin{figure*}[t]
    \centering
    \includegraphics[width=0.9\textwidth]{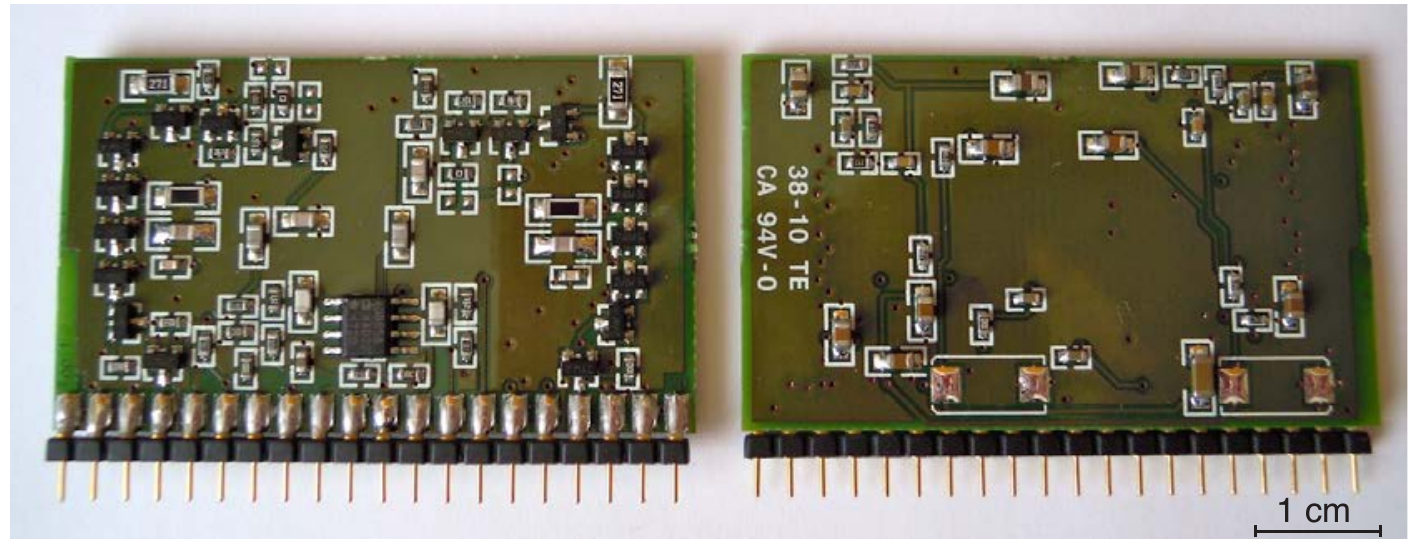}
    \caption{Photograph of two preamplifier circuits realized with
      discrete components on a four layer PCB.}
    \label{fig:preamp-picture}
  \end{figure*}

  The basic layout, shown in Figure~\ref{fig:preamp-scheme}, consists
  of a protection circuit at the input, a charge amplifier and a
  shaper followed by a non-inverting linear amplifier. 
  The circuit of the principal charge-amplifier loop with four NPX
  BF862 JFETs at the input is based on a
  preamplifier design for high-capacitance silicon
  detectors~\cite{Boiano:2004}. For the RC feedback we have chosen a
  1~pF capacitor C0 which is discharged through a resistor R0 with a
  time constant $\tau_F=470$~$\mu$s. 
  The shaping of the signal is then done with a  RC-CR
  configuration which includes a mechanism to suppress the undershoot caused
  by the feedback RC: the resistor R7 is chosen such that the product
  with the parallel capacitance C12 approaches $\tau_F$. If they are
  well matched, the zero pole caused by the feedback time constant is
  cancelled. The final transfer function of amplifying loop and RC-CR
  shaper is given by the equation~\cite{Lussi:2013}:
    \begin{equation}
    H(\omega)\propto \frac{\omega}{(i\omega\tau_D+1) (i\omega\tau_I+1) },
    \label{eq:rc-cr}
  \end{equation}
  with the computed values for the time constants $\tau_I=0.48$~$\mu$s
  and $\tau_D\approx2.75$~$\mu$s. After the shaping, the required voltage
  sensitivity of about 10~mV/fC is reached by means of an operational
  amplifier\footnote{AD8656 operational amplifier, Analog 
    devices, \url{http://www.analog.com}} in a non-inverting
  configuration with a set gain of~30. Figure~\ref{fig:preamp-picture} shows
  the realisation of two preamplifier circuits with discrete
  components on a PCB layout with four layers. The preamplifier
  response function was then explicitly derived from Equation
  (\ref{eq:rc-cr}) with an additional low pass filter (time constant
  $\approx \tau_I$) coming form the non-inverting operational
  amplifier loop after the shaper. The resulting response function is given by~\cite{Lussi:2013}:
  \begin{equation}
    V_{response}(t)=I\frac{e^{t/\tau_D}\tau_D-e^{t/\tau_I}(\tau_D+t\frac{\tau_D-\tau_I}{\tau_I})}{(\tau_D-\tau_I)^2}.
    \label{eq:response}
  \end{equation}
  
  \begin{figure*}[t]
    \centering
    \includegraphics[width=0.49\textwidth]{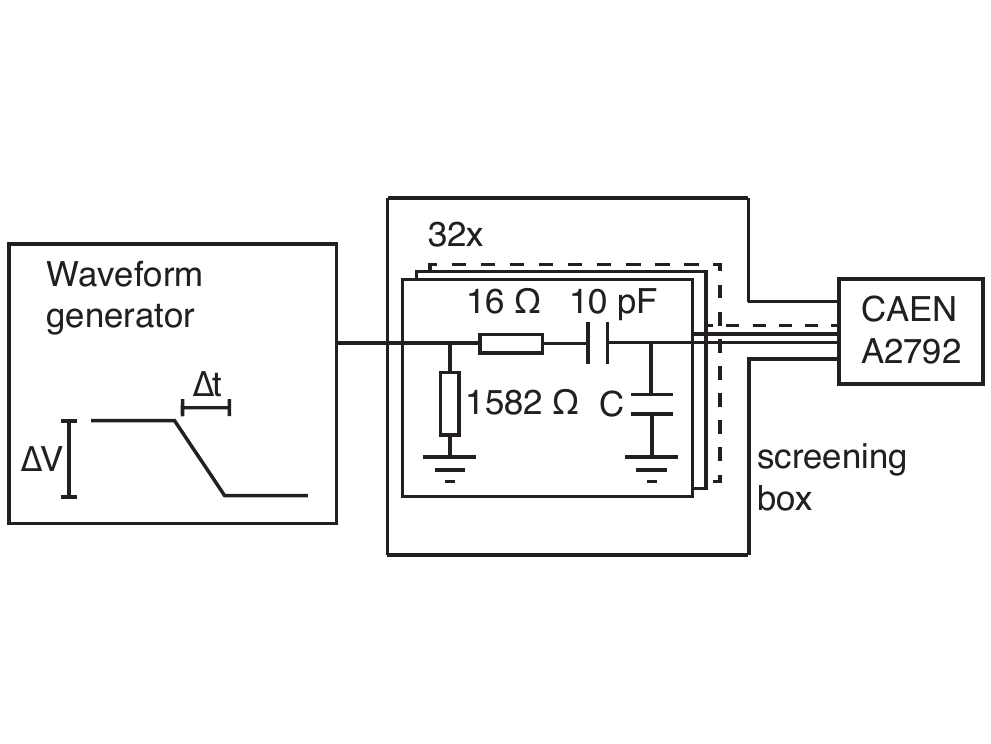}
 \includegraphics[width=0.49\textwidth]{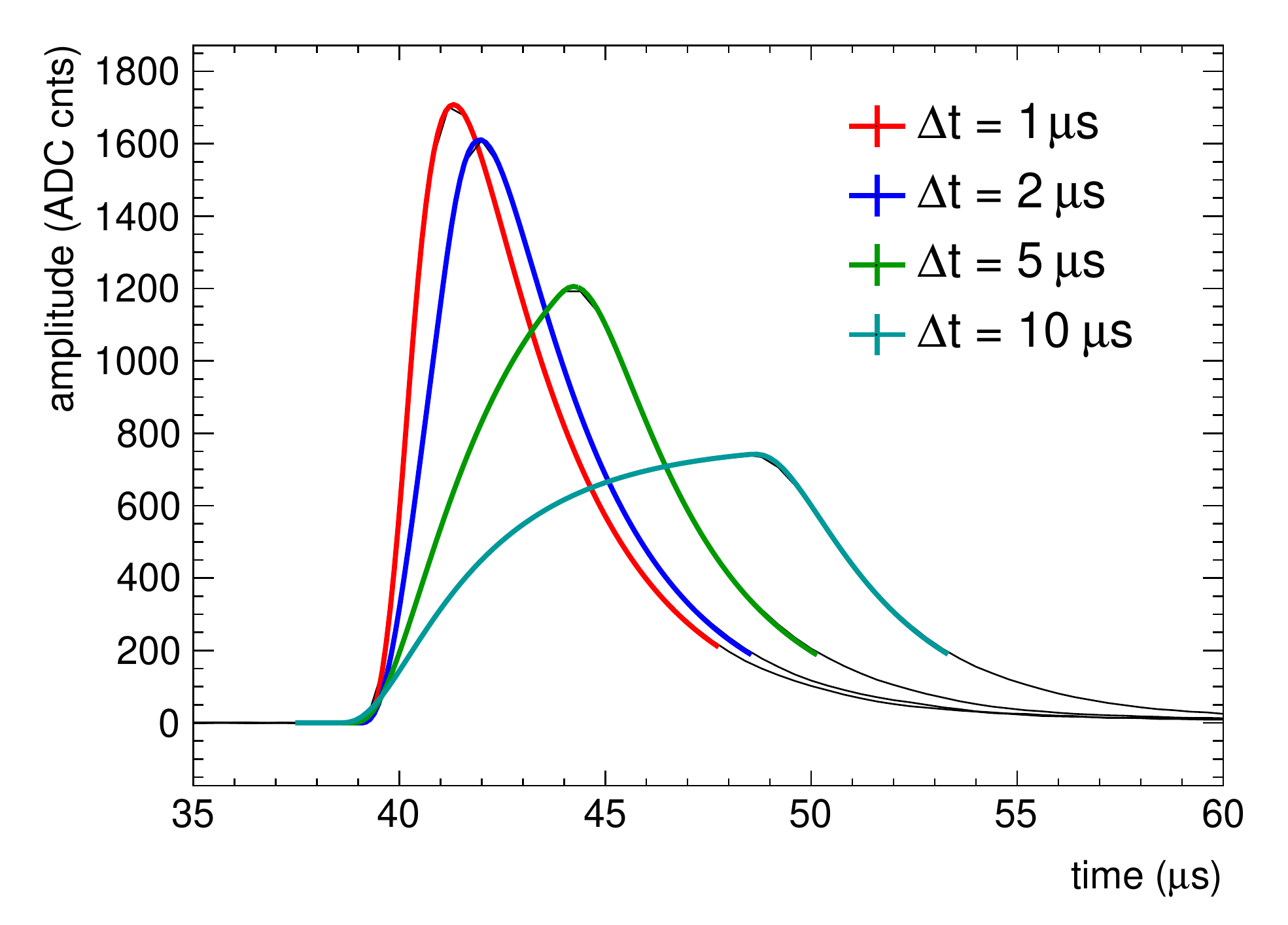}
    \caption{Left: electric scheme of the test setup for the performance
      measurements of 32 preamplifiers.
      Right: preamplifier response data (black) of a charge
      injection with different time intervals from 1~$\mu$s to
      20~$\mu$s superimposed with the corresponding fits (colored curves). [See text for details].}
    \label{fig:test-setup}
  \end{figure*}
  
  The performance of 32 preamplifiers was tested with the
  setup presented in Figure~\ref{fig:test-setup}. It allows to simultaneously feed with  a
  well defined test charge 32 preamplifiers, housed on a CAEN A2792
  acquisition board (see
  Section~\ref{sec:charge-readout-system}). A factor 1/100 attenuator
  as well as $10\pm0.25$~pF test capacitors and additional input
  capacities $C$, to simulate different detector and cable capacitances,
  are inside a screened box. The trapezoidal voltage pulse shown in 
  Figure~\ref{fig:test-setup} allows to input a constant current with
  different durations $\Delta t$ to each preamplifier. The
  voltage step $\Delta V$ was measured with a Keithley~2000
  Multimeter\footnote{Keithley Instruments Inc., \url{http://www.keithley.com}.}.  
  
  The signals shown in Figure~\ref{fig:test-setup}  were obtained
  by pulsing the test capacitor with a constant voltage step $\Delta
  V$, but with different duration $\Delta t$, thus keeping the injected charge constant.
  The fitted preamplifier response, convoluted with a constant current
  over a time $\Delta t$, is superimposed in colour in
  Figure~\ref{fig:test-setup}  to the digitised waveforms. The
  resulting time constants, given in Table~\ref{table:measurements}, agree well with the
  computed values.
 
  Besides the signal shape, we have investigated the voltage
  sensitivity and linearity by measuring the signal amplitude for
  input charges in the range from 10~fC to 180~fC. For these
  measurements, a very short pulse width of $\Delta  t=0.1$~$\mu$s was
  chosen.
  A summary of the performance measurements of 32 different
  preamplifiers is given in Table~\ref{table:measurements}. The
  RMS value of the intrinsic equivalent input noise charge (ENC) was
  measured with four different test capacities $C$ at the input of the
  preamplifiers. 
  \begin{table}[htbp]
 \caption{Average parameters measured with 32 different
      preamplifiers. The given error equals the observed spread.}
    \label{table:measurements}
    \centering
    \begin{tabular}{@{} ll@{}} 
      \hline
shaping time $\tau_D$      	&  $2.8   \pm 0.1 \hspace{2mm} \mathrm{\mu s}$	\\
      shaping time $\tau_I$      	&  $0.45 \pm 0.02 \hspace{2mm}\mathrm{\mu s}$	\\
      sensitivity                     	&  $13.8 \pm 0.4 \hspace{2mm}\mathrm{mV/fC}$		\\
      open loop gain			&  $\approx 10^4$		 					\\
      linearity (0-180 fC)		&  $\pm 1\%$ 								\\
      \begin{tabular}{@{} ll@{}}      
     
        ENC in RMS: & $C=10$~pF \\
                           & $C=92$~pF \\
                           & $C=210$~pF \\
                           & $C=480$~pF \\
      \end{tabular}
&  
     \begin{tabular}{@{} l@{}}      
       $470 \pm 30$ $e^-$ \\
       $580 \pm 30 $ $e^-$ \\
       $770 \pm 30 $ $e^-$ \\
       $1420 \pm 30 $ $e^-$ \\
     \end{tabular}
     
     \\
     \hline
    \end{tabular}
  \end{table}
  
\subsection{Data acquisition system}
\label{sec:charge-readout-system}
\begin{figure*}[t]
    \centering
    \includegraphics[width=1.0\textwidth]{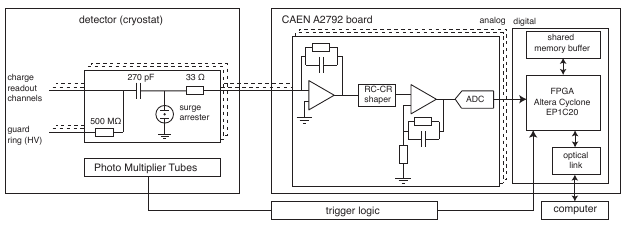}
    \caption{Overall layout of the setup with the
      detector on the left and the readout electronics on the right.}
    \label{fig:readout-scheme}
  \end{figure*}

  The overall setup shown schematically in
  Figure~\ref{fig:readout-scheme}, can be divided in two parts: the
  detector and the electronics readout crates.
  Starting from the electrodes of the 2D anode which are usually kept
  at HV to optimize the charge collection, the signals are first
  decoupled with 270~pF HV capacitors\footnote{Kekon HV capacitors, 270 pF, 10 kV, NP0,
  \url{http://www.kekon.com}}. 
  Since the LEM is operated in
  pure argon gas, it may occur that occasionally electron avalanches
  inside the LEM holes turn into streamers which short the two LEM
  electrodes, finally inducing a discharge on the anode. A multiple
  stage discharge protection was implemented to avoid damage to the
  electronics: as first stage, surge arresters\footnote{EC 90,EPCOS
    AG, Munich, Germany}, mounted close to the decoupling capacitors
  inside the detector, open to ground when voltages exceed $\sim$90 V;
  low leakage double diodes\footnote{BAV199 from NPX}, followed by a
  10$\Omega$ resistor, and a JFET are directly implemented on the
  preamplifier circuit as shown in Figure~\ref{fig:preamp-picture}. 
  \begin{figure*}[t]
    \centering
    \includegraphics[width=.8\textwidth]{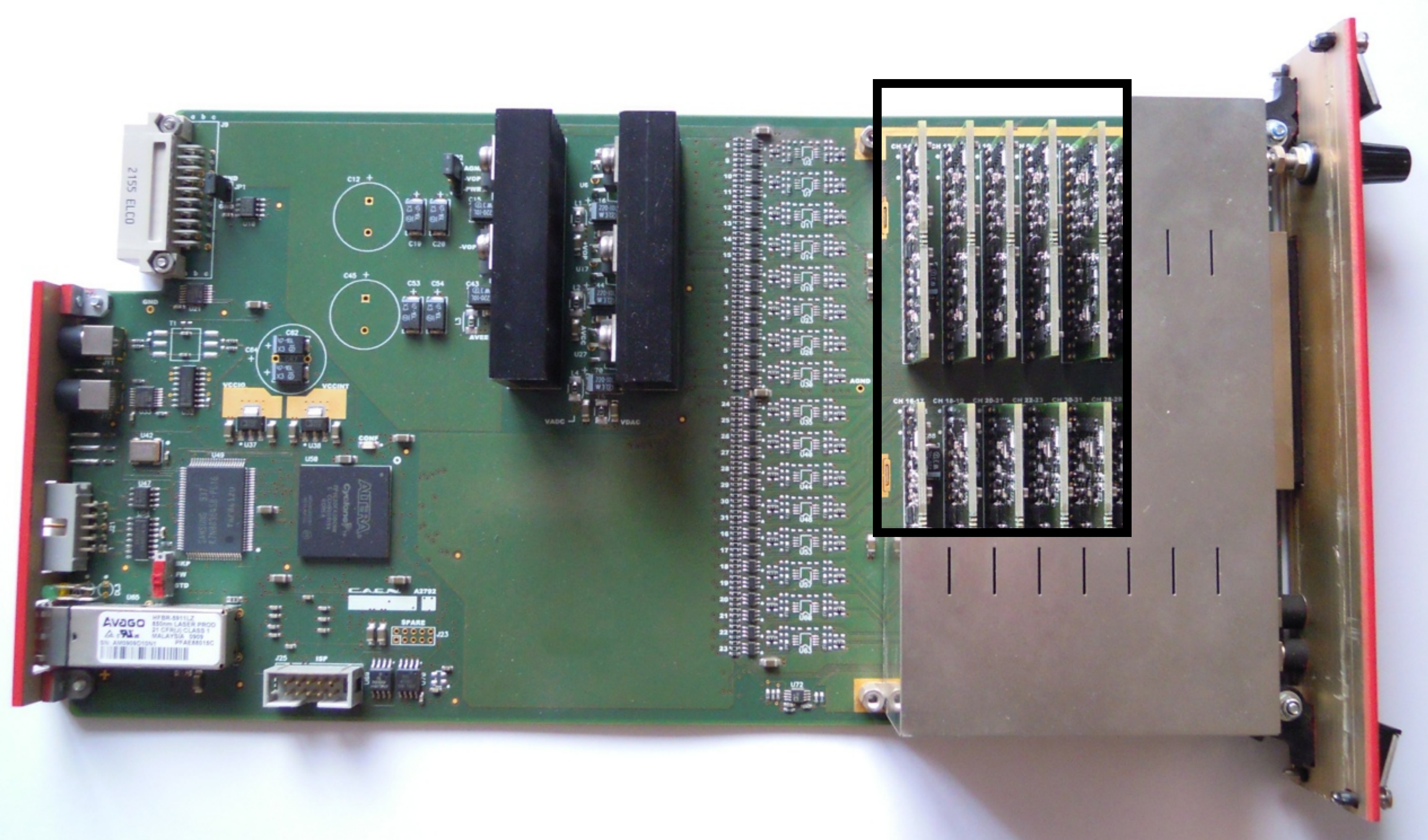}
    \caption{Picture of a CAEN A2792 readout board. The window on the
      right shows the plugged preamplifier circuits below the screen. }
    \label{fig:picture-a2792}
  \end{figure*}
  The readout system consists of several CAEN SY2791 crates
  designed and built for this project in collaboration with CAEN, each
  hosting a linear regulated power supply and 8 CAEN A2792 32 channels
  readout boards. Detector and readout system are interconnected with
  a 32 signal shielded flat cable. The preamplifiers (see
  Figure~\ref{fig:preamp-picture}) are directly plugged on the A2792
  boards, as shown in Figure~\ref{fig:picture-a2792}. In order 
  to reduce pickup noise the preamplifiers are fully surrounded by a
  screen connected to ground. A relevant feature of the design is that
  the readout boards contain both the analog and the digital sections,
  as illustrated on the right part of Figure~\ref{fig:readout-scheme}. 
  Signals are first amplified, shaped and then digitised by individual
  12 bit 2.5 MS/s ADCs with serial readout interface. The 32
  digitised signals are further processed by an FPGA, that continually
  stores the data in 1 MB circular memory buffers for each channel
  independently, it provides the trigger logic and controls the
  transfer of the data to a computer via an optical link. The time
  synchronisation between different boards and different crates is
  done with a single wire connection (TT-link) in a daisy-chain
  configuration. In addition to the clock signal, which is provided by
  the master board of a crate, also commands like {\it start}, {\it stop} and
  {\it trigger alert} are propagated through the TT-link. The system
  provides a sophisticated channel by channel trigger with two
  programmable thresholds: in case the signal exceeds a threshold value
  on any channel, a trigger alert signal is forwarded via the TT-link to
  all the other channels, causing them to lower the threshold to a
  second predefined value. Alternatively, the acquisition can also be
  triggered globally with an external signal. Such a trigger can for
  example be generated by the prompt scintillation light detection.

  \begin{figure*}[t]
    \centering
    \includegraphics[width=1.0\textwidth]{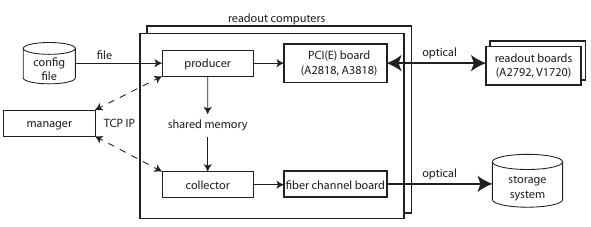}
    \caption{Schematics of the data acquisition system.}
    \label{fig:sw-scheme}
  \end{figure*}
  Figure~\ref{fig:sw-scheme} shows the schematics of the data
  acquisition system. The optical readout link can connect up to eight
  A2792 to a CAEN A2818 (PCI) or A3818 (PCIE) card on a computer. 
  
  The data acquisition system is based on commercial Linux PCs.
  In order to increase the data throughput, different crates can be
  connected to different computers. On each readout computer two
  programs are running in parallel, the \textit{producer} and the
  \textit{collector}. The \textit{producer} task is to
  send commands to the readout boards and to read the data from each
  board in event blocks. This data are then directly written to the
 \textit{shared memory}. The \textit{collector} reads from the
  shared memory and writes data to a file on an external storage
  disk. All programs (\textit{producer} and \textit{collector})
  running on different readout computers are controlled by a user
  interface (\textit{manager}). It allows the user to load a
  configuration and to start and stop the acquisition.

\section{First operation and performance of the LEM-TPC}
  \label{sec:performance}

\subsection{Detector commissioning and operation}
The filling of the detector vessel can be subdivided into three
phases: first, the main vessel volume containing
the detector was evacuated down to about $10^{-5}$~mbar
in order to let the materials outgas (the observed rate of outgassing
was estimated to be $10^{-3}$mbar l/s). Next, after adding 1~bar of pure
argon gas, the gas recirculation was turned on,
efficiently further
removing outgassed molecules. 
At this stage, the purity in the warm gas is on the order of 100~$ppb$.
At the same time the external bath was filled with LAr,
cooling down the detector. Once the system was in thermal equilibrium,
the cryostat could be filled with liquid argon obtained commercially which
typically contained $ppm$ impurities of oxygen and other elements. 
In order to reach liquid argon impurity
concentrations at the level of $ppb$, the liquid was passed through a custom made purification
cartridge which contains a stack of reduced CuO powder and 3~A molecular
sieve\footnote{Zeochem Z3-01, \url{http://www.zeochem.ch}.}.
After precise adjustment of the LAr level in the middle of the two
extraction grids, we have operated the detector during about
four weeks, starting with the commissioning of the Greinacher
circuit, the purification system and finally also the LEM-TPC. 

The signals of the detector were read out with the acquisition system described in
Section~\ref{sec:charge-readout-system}. The 2D anode with
2$\times256$ readout strips required the use of two fully equipped
CAEN SY2791 crates. It was globally triggered by the prompt
scintillation light, measured with the two cryogenic PMTs positioned
below the cathode grid of the detector. With a 20~Hz trigger rate we
achieved a stable data throughput of about 40~MB/s. 

\begin{figure*}[t!]
\centering
\includegraphics[width=0.9\textwidth]{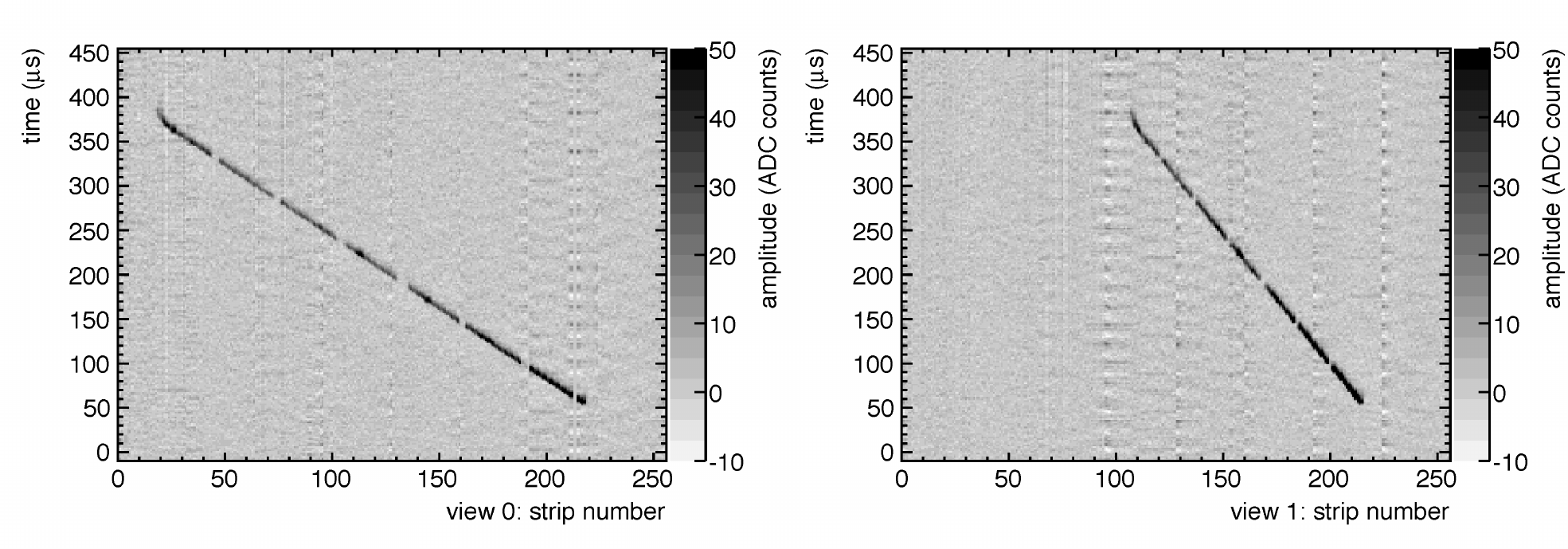}
\includegraphics[width=0.9\textwidth]{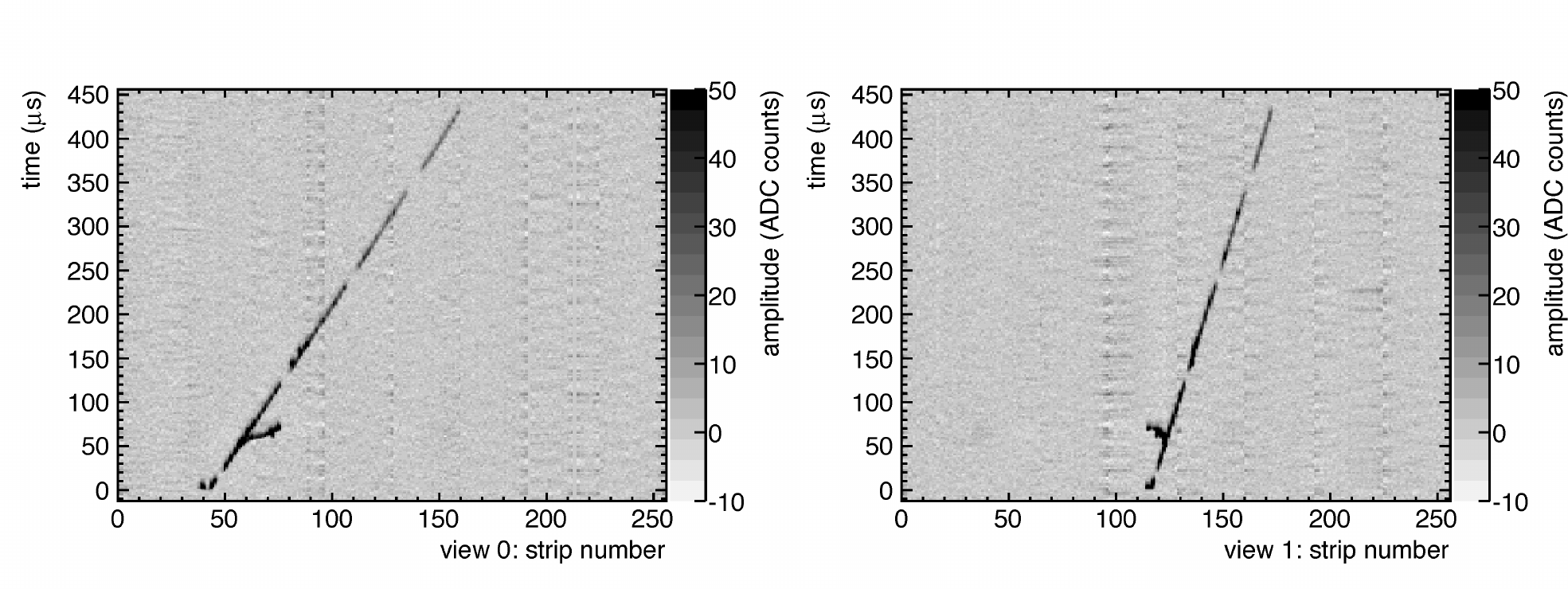}
\includegraphics[width=0.9\textwidth]{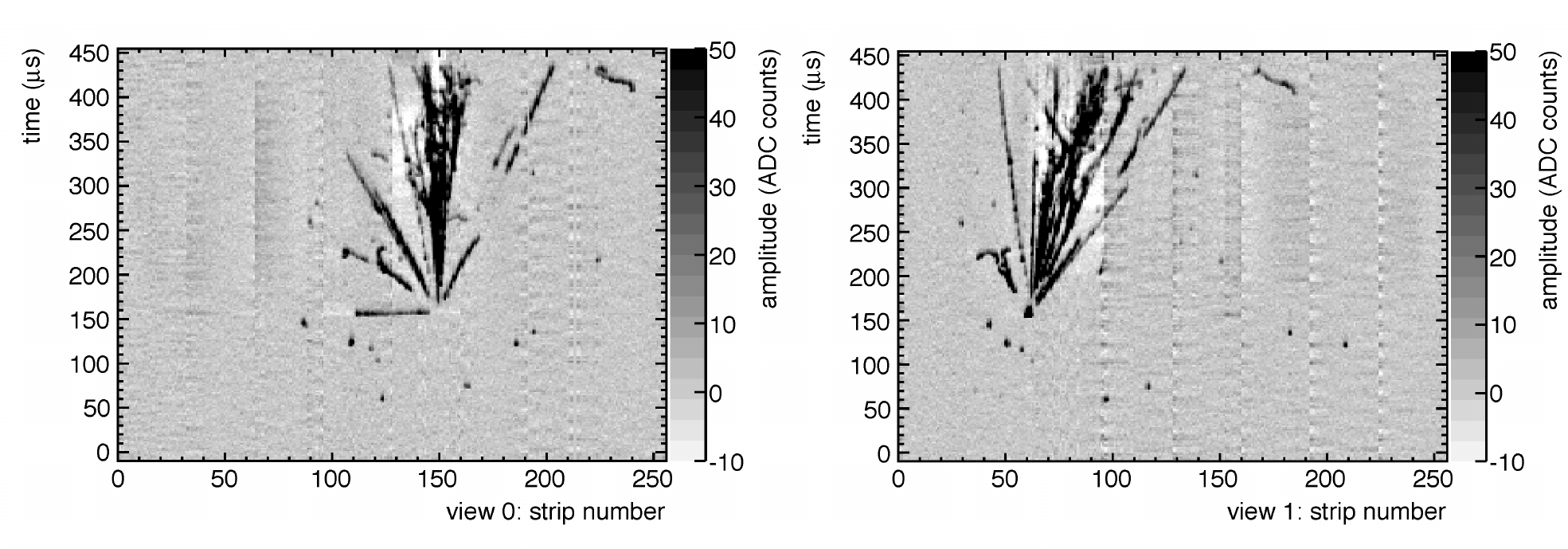}
\includegraphics[width=0.9\textwidth]{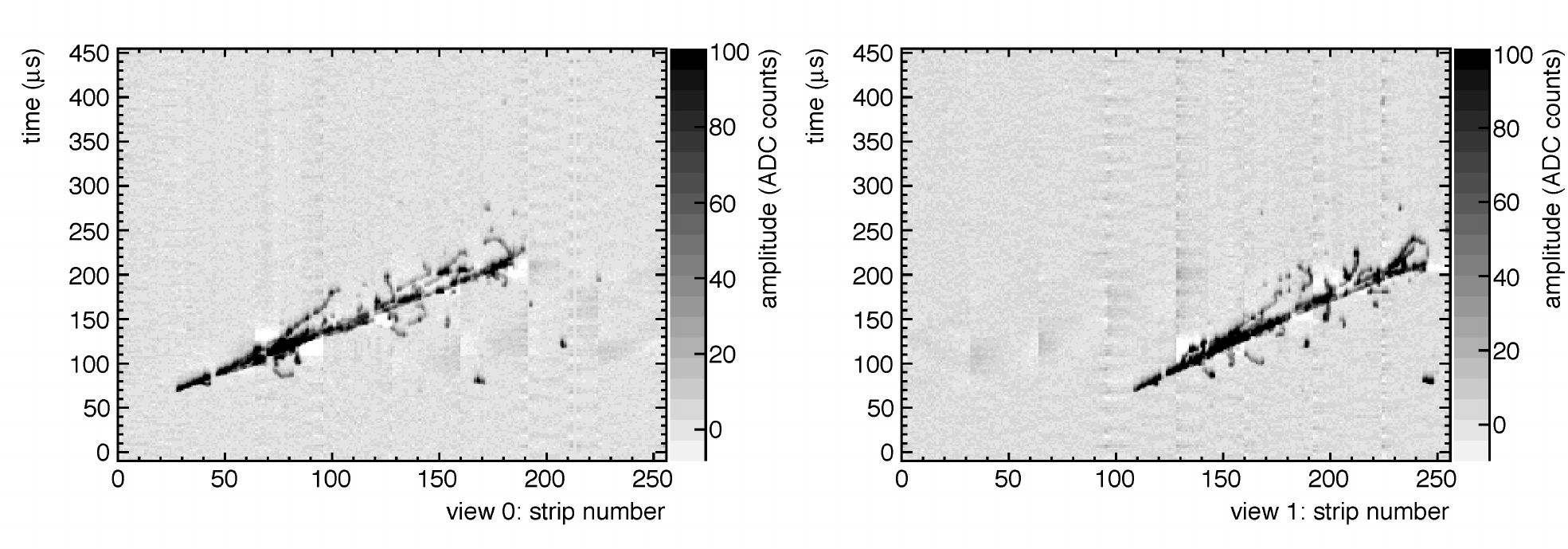}
\caption{Cosmic ray event displays (left: view 0, right view 1),
recorded in double phase conditions with a LEM
  field of 35~kV/cm. The four event displays show from top to bottom
  two muons, a hadronic and an electromagnetic shower candidate. [See
  text for more details.]} 
\label{figure:eventGallery}
\end{figure*}

Stable operation of the detector was reached with the nominal electric
fields reported in Table~\ref{table:FieldConfiguration}. 
\begin{table}[t]
\centering
\begin{tabular}{ l l l }
\hline
                                & dimension   & electric field \\
\hline
anode-LEM		& 0.2~cm        & 2~kV/cm\\
LEM          		& 0.1~cm        & 30--35~kV/cm\\
grid-LEM			& 1.2~cm        & 0.6~kV/cm\\
extraction	        & 1.0~cm        & 2~kV/cm (in LAr)\\
drift				& 60~cm         & 0.4~kV/cm\\
\hline
\end{tabular}
\caption{Nominal electric field configuration and dimensions of the
  LEM-TPC. The absolute potential of the anode was kept at +1~kV.}
\label{table:FieldConfiguration}
\end{table}
Figure~\ref{figure:eventGallery} shows four typical cosmic ray events, taken from
the highest gain run with an amplification field of 35~kV/cm. From top
to bottom, there are two cosmic muons crossing the detector, a
deep inelastic interaction and an electromagnetic shower candidate. 
These plots show the usual representation of events with the drift 
time as a function of the strip number for both views. The greyscale
represents the signal amplitude for each sample in a linear scale.
It can clearly be seen that both views show symmetric unipolar signals
which are clearly distinguishable from the noise. As a consequence of
the good signal to noise ratio and the spatial resolution, small
structures like knock-on electrons -- also called $\delta$-rays -- can be identified and
reconstructed in three dimensions as demonstrated below. Since cosmic muons
produce straight crossing tracks and the energy deposition is known to
be $\sim$2.1~MeV/cm, the events can be used to characterise the detector
in terms of free electron lifetime and amplification. An obvious limitation of
this charge readout was the inactive area introduced by the
segmentation of the LEM. The 1.6~mm wide gaps between the LEM
electrode segments appear as equally spaced missing charge, as seen in
the two muon events in Figure~\ref{figure:eventGallery}.

\subsection{Event reconstruction}
The event reconstruction of the digitised raw waveforms
was done with the so-called Qscan software package for LAr
detectors~\cite{Rico:2002}. Here we give a brief description of the basic
steps in the reconstruction procedure. A more detailed description of the
event reconstruction in a LAr-TPC can be found
elsewhere~\cite{Lussi:2013,Rico:2002}. Due to the presence of 
coherent pick up noise at our laboratory location, which could not be eliminated during the run,
the events had to be digitally filtered. As explained
in Ref.~\cite{Badertscher:2011sy}, an algorithm to subtract common
noise on all channels has been applied. This algorithm basically makes
use of the fact that the same coherent noise is picked up by
all the electrodes and subtracts it. In addition, remaining
discrete noise frequencies, which appeared dominantly on all the
channels in the power spectrum, have been suppressed. Finally, a
pedestal value, which is computed as the average amplitude in the pre-trigger
window from 0 to 56~$\mu$s, is subtracted from each waveform.

After the digital signal processing of the waveforms the so called
hit-finding algorithm discriminates physical signals from noise. The
physical parameters which can be extracted from found hits are
the time when the signal rises and the hit integral. Since the events
are always triggered by the prompt scintillation light, the signal time
is equal to the drift time of the charge. The methods to extract
the drift velocity and its uniformity has been reported
elsewhere~\cite{Badertscher:2012dq} and found to be consistent with $v_d \sim 1.4$~mm/$\mu$s.
The hit integral, after
calibration, is a direct measurement of the ionisation charge $\Delta Q$. After
clustering adjacent hits, straight tracks have to be found and
reconstructed. This is done by converting the hit parameters from the
real space (strip and drift time) to the parameter space of straight lines.
In the Hough space~\cite{Hough}, lines appear as points, thus the problem
of finding hits in a straight line is reduced to the problem of
finding maxima in the parameter space of straight lines. Another
advantage of the Hough algorithm is that it is not sensitive to
missing hits due to the LEM segmentation, as discussed above. 
\begin{figure*}[t]
\centering
\includegraphics[width=1.0\textwidth]{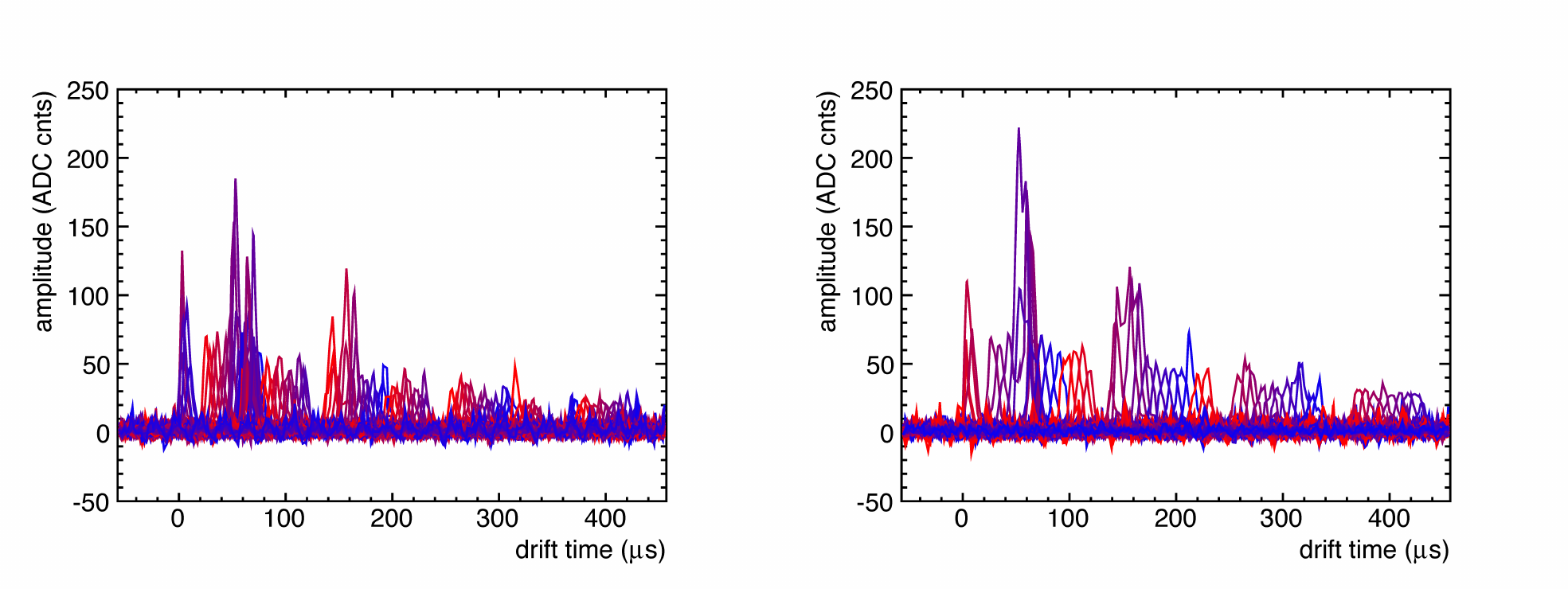}
\includegraphics[width=1.0\textwidth]{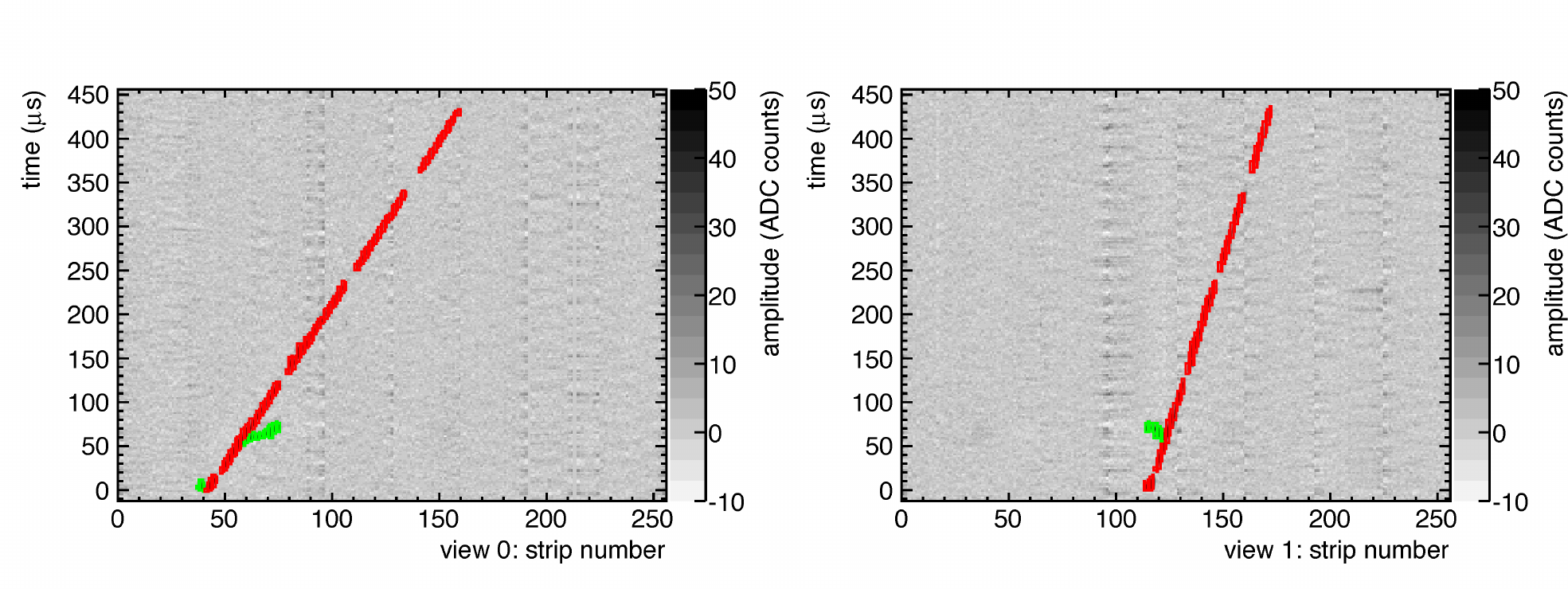}
\caption{Typical cosmic muon crossing the full drift length of 60 cm. 
  Top: signal waveforms of all the channels of the two views; bottom:
  event display representation of the same event showing drift time versus anode strip number. Found hits of the main $\mu$
  track are marked in red, identified $\delta$-ray hits are marked in green.}
\label{fig:event-14285-2D}
\end{figure*}
In the event display shown in Figure~\ref{fig:event-14285-2D} all the
hits which are identified as part of a straight line are marked in
red. In order also to reconstruct $\delta$-rays with enough energy to
produce secondary tracks (few MeV) an algorithm that searches for residual hits
which are attached to but not part of the main muon track was implemented. If such an
attachment of the main track (green) consists of at least two hits, it
is identified as $\delta$-ray. 

\begin{figure*}[t]
  \centering
  \includegraphics[width=0.7\textwidth]{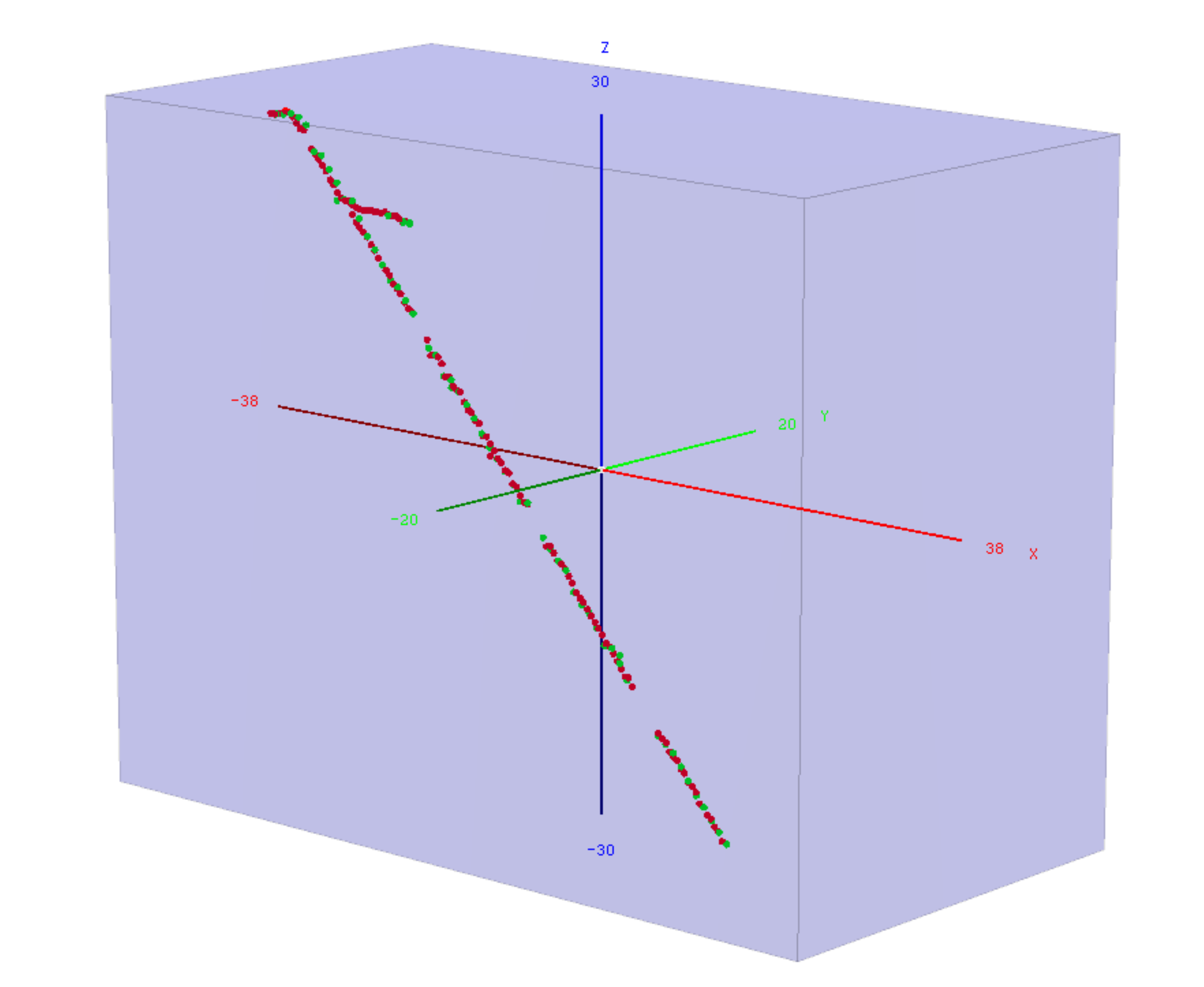}
  \caption{Crossing muon event with a $\delta$-ray
    reconstructed in three dimensions. Each dot corresponds to a three
    dimensional hit of the track. View~0 (1) hits are shown in red
    (green).}
  \label{fig:event-14285-3D}
\end{figure*}
After matching two main tracks with equal time endpoints, the three
dimensional coordinates of its hits can easily be computed as well as
the three dimensional track pitch $\Delta x$ of each hit. Together
with the information of its charge $\Delta Q$ the charge per unit length $\Delta
Q/\Delta x \approx dQ/dx$ can be computed. 
Since $\delta$-rays are typically not
straight as it is the case for the cosmic muons, each $\delta$-ray hit is matched by
time with a $\delta$-ray hit of the second view. In case the $\delta$-ray
appears only in one view, the hits are matched with the main muon
track. Figure~\ref{fig:event-14285-3D} shows a three-dimensionally
reconstructed muon track with a $\delta$-ray. 

\subsection{Free electron lifetime measurement}
As described in Section~\ref{sec:detector} the cryogenic system was
equipped with two separate purification systems. During the run period
both systems were operated, thus it was important to monitor the
liquid argon purity. In order to measure the charge attenuation as a
function of the drift time, we have used the reconstructed cosmic muon
tracks.  
\begin{figure*}[t]
  \centering
  \includegraphics[width=0.49\textwidth]{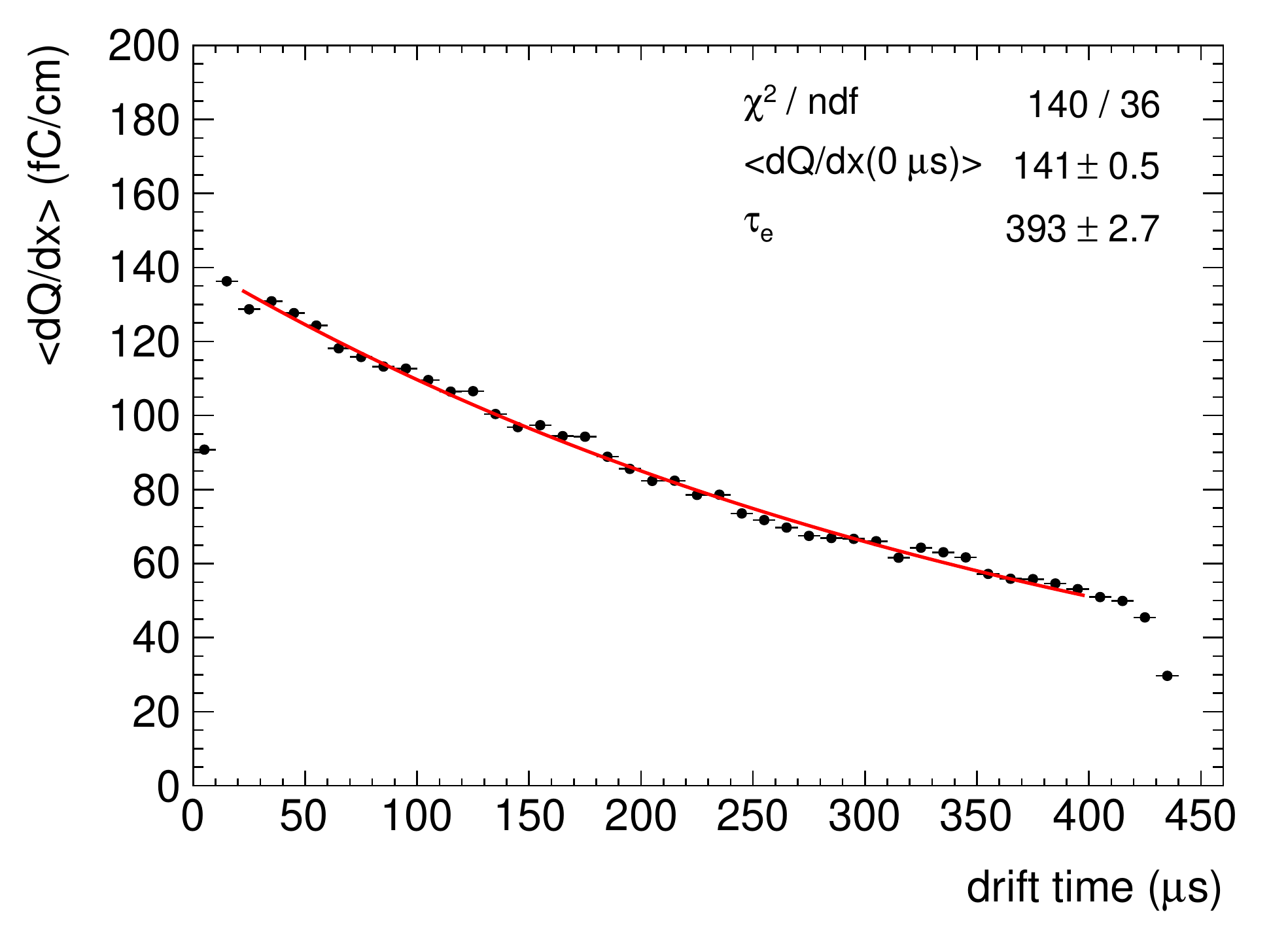}
  \includegraphics[width=0.49\textwidth]{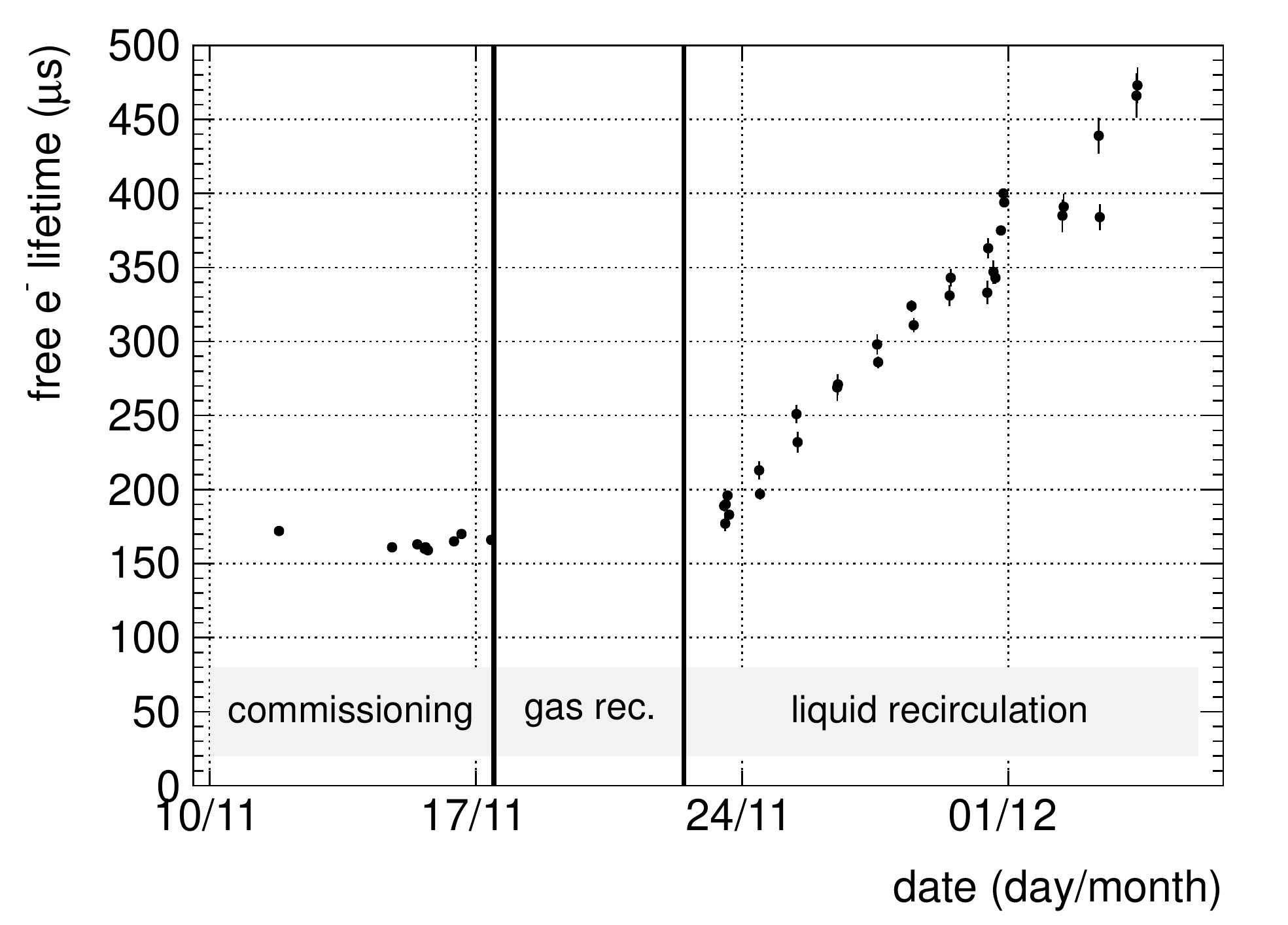}
  \caption{Left: $<dQ/dx>$ as a function of the drift time
    superimposed with an exponential function with the decay time constant
    $\tau_e=393\pm3$~$\mu$s fitted to the data points.  Right:
    evolution of the free electron lifetime during the run period.} 
  \label{fig:lifetime}
\end{figure*}
The data points in Figure~\ref{fig:lifetime} show the mean $dQ/dx$
values of all reconstructed muons for different drift times. It can be
seen that the exponential function $e^{-t_{drift}/\tau_e}$ with the free electron
lifetime $\tau_e$ perfectly fits the obtained distribution. 
The maximum electron drift time is about 430~$\mu$s.
The plot
on the right side shows the lifetime 
evolution during the last three weeks of operation. Starting from a
lifetime of about 170~$\mu$s we first did initial tests with the
liquid recirculation. During this commissioning period the free electron lifetime
remained constant. In a second phase, the gas recirculation was turned
on. Due to changes in pressure and liquid argon level we did not
acquire data with the double phase readout. However, after stopping
the gas recirculation, the measured free electron lifetime was not
significantly improved although was maintained at constant value. 
During
the last phase we have  operated the liquid argon recirculation
system at a flow rate of about 5 to 7~lt LAr/h. 
Since we saw a steady increase of the
free electron lifetime until we stopped the run, we conclude that the liquid argon
recirculation system was efficiently working. When we stopped the run
a lifetime of 470~$\mu$s was measured. This corresponds to an oxygen
equivalent impurity concentration of about 0.64~ppb. 

\subsection{Amplification gain and energy spectrum of $\delta$-rays}
After the determination of the free electron lifetime, the factor
$e^{t_{drift}/\tau_e}$ has to be used to correct for charge losses due
to the electron attachment to impurities. This allows then to do a
measurement of the effective gain and the signal to noise ratio for
the recorded MIP events. As explained in
detail in reference~\cite{Badertscher:2011sy} the effective gain of
the device is defined as the ratio of the measured charge collected 
(and corrected for the measured electron livetime) on  
 both views $<dQ_0/dx>+<dQ_1/dx>$ and the initially produced
charge $<dQ/dx>_{MIP}=10$~fC/cm. This means that this factor includes
the intrinsic Townsend avalanche as well as losses at the liquid-gas
interface, at the top extraction grid and at the LEM.   
\begin{figure*}[t]
  \centering
  \includegraphics[width=0.49\textwidth]{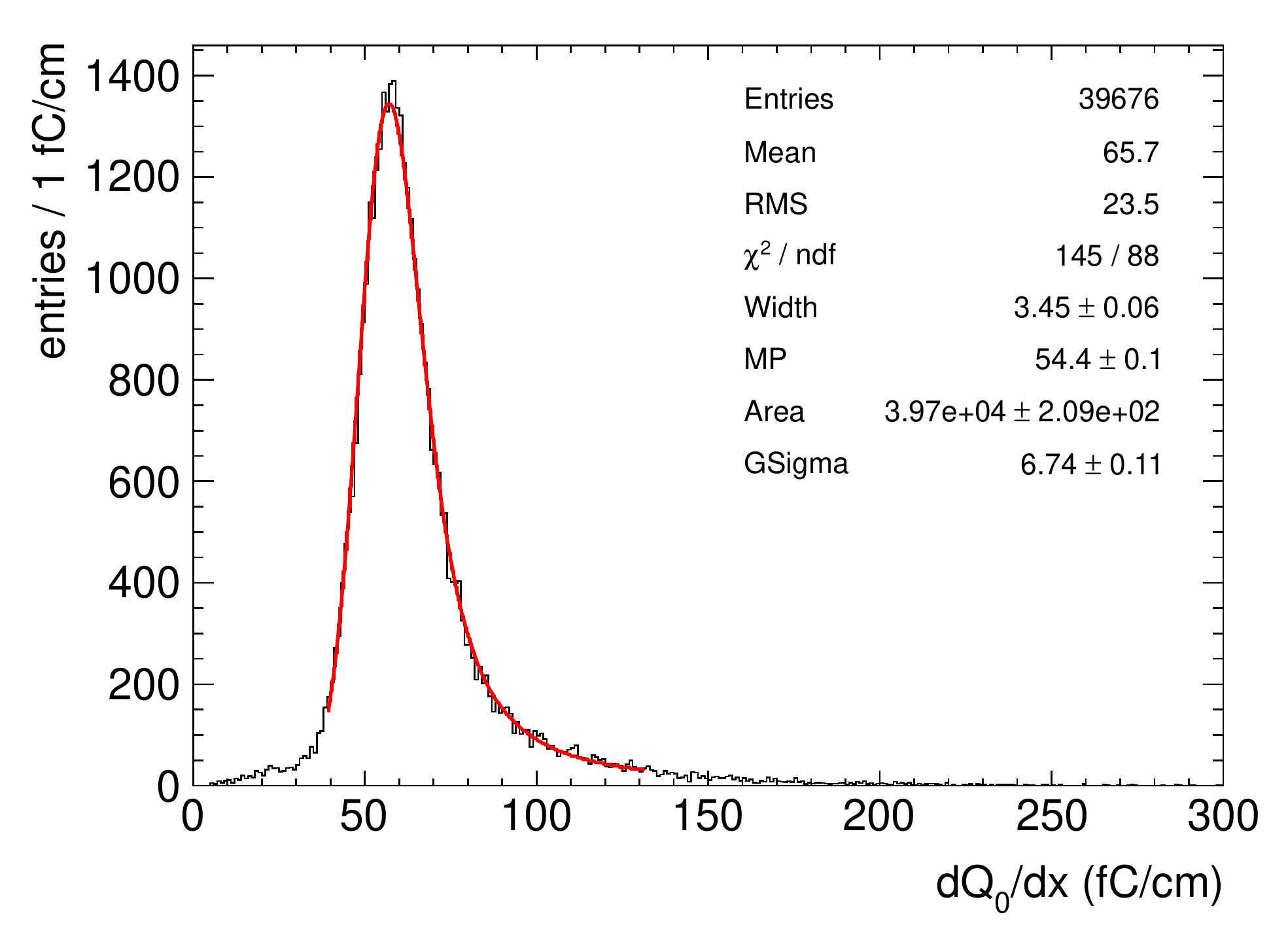}
  \includegraphics[width=0.49\textwidth]{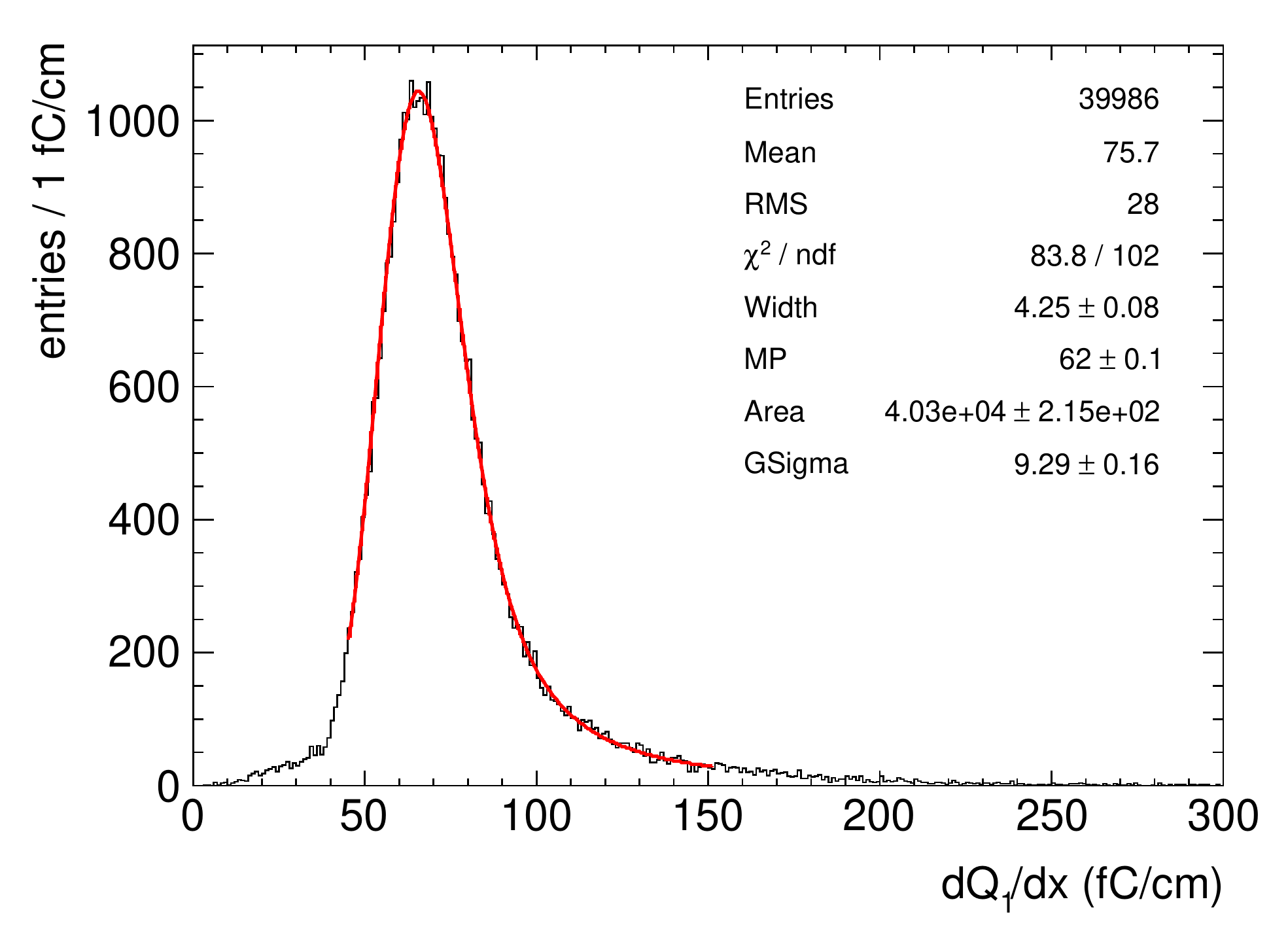}
  \caption{dQ/dx distribution for both views for selected straight tracks from a
    run with an amplification field of 3.5~kV across 1~mm. A Landau function, convoluted with a Gaussian is fitted to the distributions.}
  \label{fig:dqdx}
\end{figure*}
The two plots in Figure~\ref{fig:dqdx} show the 
$dQ/dx$ distributions for view~0 and view~1 from the run with 35~kV/cm
amplification field, superimposed with a fitted Landau convoluted
with a Gaussian. Besides the
effective gain measurements it was important to verify that the strips of each
view collect the same amount of charge. We use  the measured
$dQ/dx$ distributions of both views to compute the asymmetry factor 
\begin{equation}
\frac{<dQ_1/dx>-<dQ_0/dx>}{<dQ_1/dx>+<dQ_0/dx>}\approx7\%. 
\end{equation}
In addition to a gain of 14 and a good sharing of the charge among the
two views, we also measured an excellent signal to noise ratio of
about~30 for a MIP. To compute this ratio, we used the mean amplitude of cosmic
muon induced signals divided by the average noise, defined as the
average RMS value for all the readout channels. 

\begin{figure*}[t]
  \centering
  \includegraphics[width=0.8\textwidth]{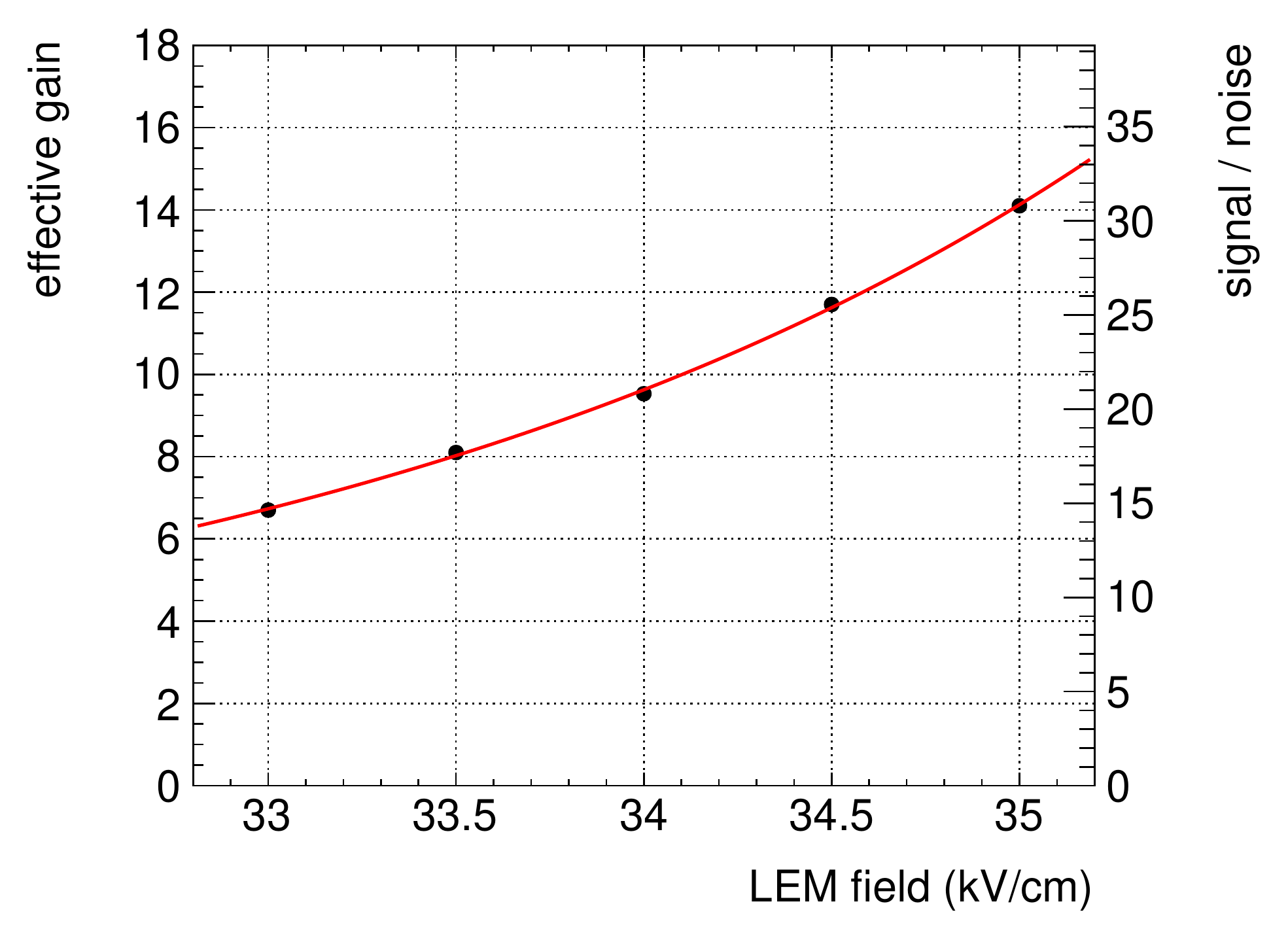}
  \caption{Effective gain of the detector (left) as a function
    of the LEM amplification field.}
  \label{fig:gain}
\end{figure*}
Figure~\ref{fig:gain} shows a summary of the effective gains obtained
with different amplification fields between 33~kV/cm and 35~kV/cm. The
values of all the other fields, defined in
Table~\ref{table:FieldConfiguration}, were kept constant. 
\begin{figure*}[t]
  \centering
  \includegraphics[width=0.9\textwidth]{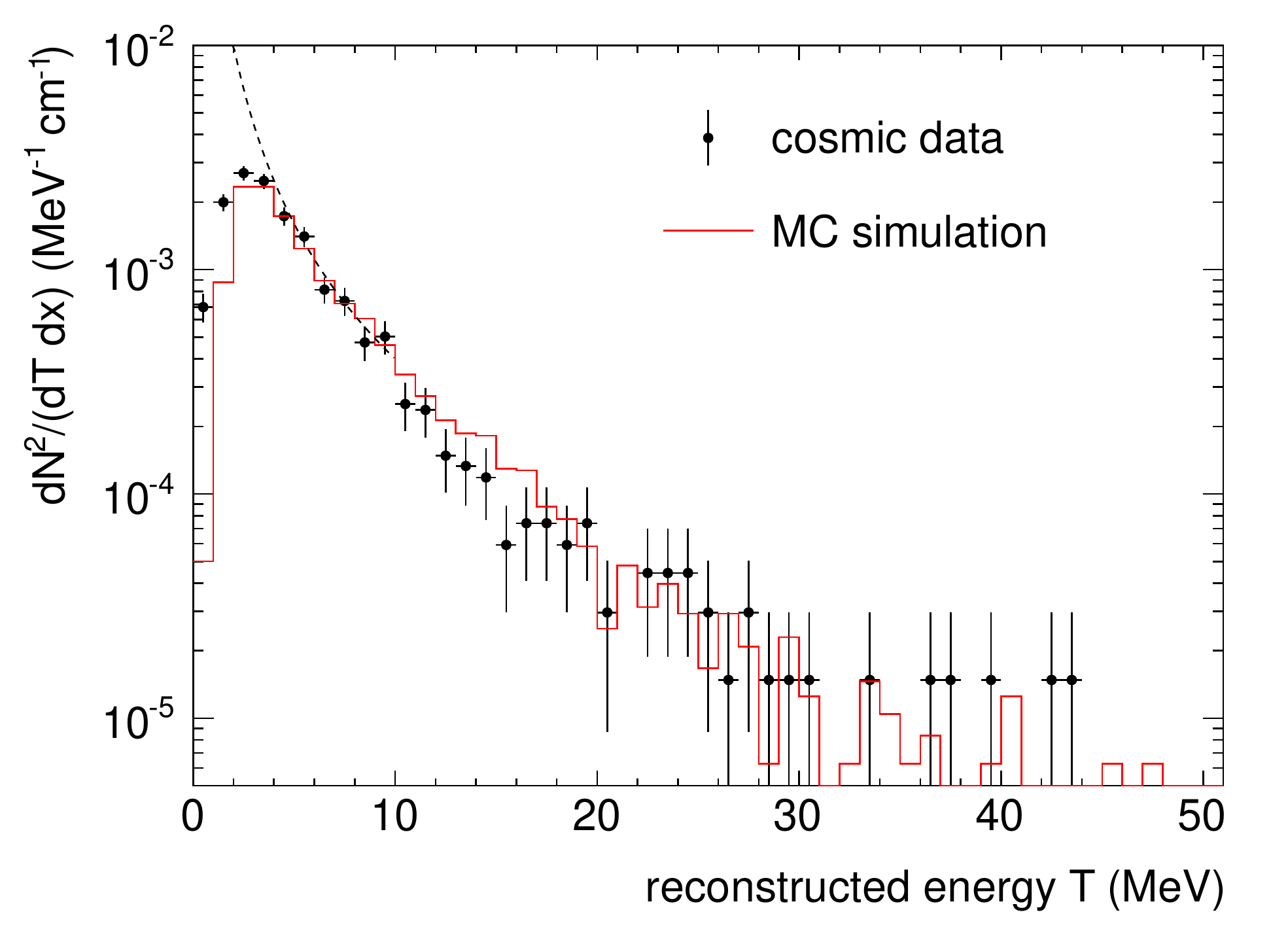}
  \caption{The reconstructed energy of $\delta$-rays that are longer
    than 0.5~cm from the collected cosmic muon sample, as well as from the MC 
    generated muon events with a momentum uniformly distributed
    between 1 and 10 GeV/c.}
  \label{fig:dray}
\end{figure*} 

Figure~\ref{fig:dray} shows the number of reconstructed $\delta$-rays, normalised by
the muon track length, as a function of the reconstructed kinetic
energy. To get an estimation of the kinetic energy, the calorimetric
approach of summing up the deposited energies of all the tagged hits
was used. Since the initial part of the $\delta$-ray is less easily
 separated from the main muon track, the obtained kinetic energy
is always an underestimation of the true value. In order to do a consistency
check, Qscan has been interfaced with Geant~4 using the Virtual Monte
Carlo package VMC~\cite{Hrivnacova:2008zza}. 
Muons with a uniform distribution between 1 and 10~GeV/c and with the 
angular distribution of the observed cosmic rays  were propagated in a 
simplified detector geometry, digitised and reconstructed, together 
with their $\delta$-rays. The obtained energy distribution of the MC 
generated $\delta$-rays is shown in Figure~\ref{fig:dray}, above 3 MeV it 
agrees well with the data from cosmic rays. 
The theoretical curve
${d^2N}/{(dTdx)}\propto {\beta^{-2}} (1-\beta^2\frac{T}{T_{max}}){T^{-2}}$,
where $T$ is the kinetic energy of the $\delta$-ray, $\beta$ its velocity,
and $T_{max}$ is defined in Ref.~\cite{pdg2012},
is also shown as dashed line below 10~MeV, showing that with the present 
algorithm, the  efficiency for reconstructing
the $\delta$-rays drops below 5~MeV.
The discrepancy between data and MC simulations
below 3~MeV shows that the efficiency for reconstruction
is  slightly under-estimated in the simulation.

\section{Conclusions}
\label{sec:conclusions}
We have produced and successfully operated
the largest LAr LEM-TPC with 2D readout anode of an area
of $40\times76$~cm$^2$ ($\sim 0.5$~m$^2$) and 
60~cm drift. During a  very successful run
with double phase ultra-pure argon at 87~K, the detector
was  exposed to cosmic rays and recorded a large number
of events during a long-term data-taking period.
A stable effective gain~14 was reached,
giving an excellent signal to noise ratio of $>30$ for a MIP.
The detector performance has been studied and a sample of $\delta$-rays 
was measured and compared to a MC simulation.
The exact performance with a gain of $\sim$30
reached with the  $10\times 10$~cm$^2$ area chamber~\cite{Badertscher:2011sy}
could not be reproduced in this first test of a detector of $\sim$0.5~m$^2$-scale.
The observed limitations such as this apparently lower maximum gain,
the introduced dead space by the segmentation of the LEM, the large
anode capacitance and the long signal cables, are presently being addressed and
will be further treated and hopefully corrected in the near future.

\acknowledgments
This work was supported by ETH Zurich and the Swiss National Science
Foundation (SNF).We are grateful to CERN
for their hospitality and thank R. Oliveira and the TS/DEM group,
where several of the components of our detector were manufactured. We
also thank the RD51 Collaboration for useful discussions and
suggestions. Special thanks go to N. Bourgeois and the CERN PLC
support group for the PLC system and to T. Schneider and Thin Film \&
Glass group of CERN for helping us for the WLS coating of the PMTs. We
thank Maria de Prado (CIEMAT), Takasumi Maruyama (KEK/IPNS), Junji
Naganoma (Waseda), Hayato Okamoto (Waseda) and 
A.~Marchionni (FNAL) for their participation in the data-taking phase.

\end{document}